\documentclass[12pt]{article}
\usepackage{psfig}

\def\lesssim{\mathrel{\mathpalette\vereq<}}
\def\gtrsim{\mathrel{\mathpalette\vereq>}}
\makeatletter
\def\vereq#1#2{\lower3pt\vbox{\baselineskip1.5pt \lineskip1.5pt
\ialign{$\m@th#1\hfill##\hfil$\crcr#2\crcr\sim\crcr}}}
\makeatother

\begin{document}

\begin{titlepage}
\begin{center}
May 4, 1999     \hfill    LBNL-43104 \\
~{} \hfill UCB-PTH-99/15  \\
~{} \hfill hep-ph/9904399\\

\vskip .1in

{\large \bf Seasonal Variations of the ${^7}$Be Solar Neutrino Flux}%
\footnote{This work was supported in part by the U.S. 
Department of Energy under Contracts DE-AC03-76SF00098, in part by the 
National Science Foundation under grant PHY-95-14797.  HM was also 
supported by the Alfred P. Sloan Foundation and AdG by CNPq (Brazil).}

\vskip 0.3in

Andr\'e de Gouv\^ea, Alexander Friedland, and Hitoshi Murayama

\vskip 0.05in

{\em Theoretical Physics Group\\
     Ernest Orlando Lawrence Berkeley National Laboratory\\
     University of California, Berkeley, California 94720}

\vskip 0.05in

and

\vskip 0.05in

{\em Department of Physics\\
     University of California, Berkeley, California 94720}

\end{center}

\vskip .1in

\begin{abstract}
Measuring the ${^7}$Be solar neutrino flux is crucial towards
solving the solar neutrino puzzle. The Borexino experiment, and
possibly the KamLAND experiment, will be capable of studying the  
$^7$Be neutrinos
in the near future. We discuss (1) how the seasonal variation of the
Borexino and KamLAND data can be used to measure the $^7$Be solar
neutrino flux in a background independent way and (2) how anomalous
seasonal variations might be used to discover vacuum neutrino
oscillations, independent of the solar model and the measurement of
the background. In particular, we find that, after three years of 
Borexino or KamLAND running, vacuum neutrino oscillations
can be either established or excluded for almost all values of 
$(\sin^22\theta,\Delta m^2)$ preferred by the Homestake, GALLEX, SAGE, and
Super-Kamiokande data. We also discuss how well
seasonal variations of the data can be used to measure 
$(\sin^22\theta,\Delta m^2)$ in the case of vacuum oscillations.   
\end{abstract}

\end{titlepage}

\newpage
\setcounter{footnote}{0}
\section{Introduction}

The question whether neutrinos have non-zero mass has been an
outstanding issue in particle physics for many decades.
Recently there have been new exciting developments in the indirect
search for neutrino masses via neutrino oscillations. Major progress
has been achieved by studying atmospheric neutrinos,
culminating in the announcement of evidence for muon neutrino
oscillations by the Super-Kamiokande collaboration
\cite{atmospheric}. The most 
striking signal presented in \cite{atmospheric} 
is the up-down asymmetry of the  
atmospheric muon neutrino flux. The
choice of this particular quantity eliminates many theoretical
uncertainties and the final result is very robust. 
In fact, at present time, this result represents perhaps the best
evidence for physics beyond the Standard Model.

Another very active area of research is the study of neutrinos
coming from the Sun. Ever since the Homestake experiment \cite{Cl} 
reported its first
results, there has been disagreement between theoretical predictions
and measurements of the solar neutrino flux. 
For many years, however, it was not possible to determine if the
observed discrepancy was due to problems with the experiment and/or
with the modeling of the Sun, or if it was, in fact, a sign of new physics. 
In the last decade other neutrino experiments, 
Kamiokande \cite{Kamiokande}, GALLEX \cite{GALLEX}, SAGE \cite{SAGE}, 
and more recently Super-Kamiokande \cite{Super-K}, 
have also measured the solar neutrino flux, with different
energy thresholds and using very different techniques. All four experiments
confirm a deficit in the observed number of solar neutrino induced
events. Moreover, it has recently become clear that it is virtually 
impossible to
concoct a solar model which would fit all the data
\cite{bksreview,ssm_indep}. 
On the other hand, the results of all experiments can be explained by 
assuming that the electron neutrino oscillates into a different flavor state.

There are two neutrino oscillation scenarios that are capable of
faithfully explaining the solar neutrino data \cite{bksreview}.  
One scenario makes use of
the MSW effect \cite{W,MS}, where the electron neutrino conversion into another
neutrino flavor is due to flavor dependent interactions with solar
matter. The other scenario assumes that the neutrino oscillation
length is comparable to the Earth-Sun distance, and simple
vacuum oscillations are sufficient. This scenario is also known as the 
``just-so'' solution \cite{glashowkrauss}. Both scenarios allow the electron
neutrinos to oscillate into other active species or sterile neutrinos.

The solar neutrino energy spectrum is determined by several nuclear
reactions which take place in the Sun's core \cite{reactions}, and
different experiments are sensitive to neutrinos produced by different
nuclear reactions. 
Super-Kamiokande, for example, a very large water Cherenkov
detector, is currently sensitive to solar neutrinos with energies
slightly above 5.5~MeV. 
Almost all neutrinos it detects
come from the decay of $^8$B ($^8$B $\rightarrow$ $^8$Be$^{*} +
e^++\nu_e$).\footnote {There is a small fraction of the neutrinos
  that can be detected at Super-Kamiokande coming from the reaction     
  $^3$He $+ p\rightarrow ^4$He $+e^+ + \nu_e$.} 

Another solar reaction that gives rise to neutrinos is the process of
electron capture
by a $^7$Be nucleus ($^7$Be $+e^-\rightarrow$ $^7$Li
$+\nu_e$). Neutrinos from this reaction have energies below the
Super-Kamiokande threshold, but are accessible to the radiochemical
experiments Homestake, GALLEX, and SAGE. If one naively assumes that
the suppression in the neutrino flux is due to the suppression of
individual neutrino sources ($^8$B, $^7$Be, etc) in the Sun, the 
combination of the Super-Kamiokande data with that of the radiochemical 
experiments indicates that the flux of $^7$Be neutrinos is virtually
absent \cite{ssm_indep,noroom47Be} (the best fit value
of the $^7$Be flux is in fact negative!). In the case of neutrino
oscillations, all solutions to the solar neutrino puzzle indicate 
that the $^7$Be neutrino flux is suppressed, in some cases very
strongly. 
Thus, at present, there is great demand for
experiments that would accurately measure the flux of the $^7$Be
neutrinos. Two upcoming experiments, Borexino and KamLAND, may have
the capability to do exactly that. 

In this paper, we present a quantitative study of what can be
accomplished by measuring the seasonal variations of the $^7$Be  
neutrino flux at Borexino and KamLAND. 
Seasonal variations of the solar neutrino flux are of course expected,
due to the Earth's eccentric orbit. The number of
neutrinos of all flavors reaching the Earth is larger when the Earth is
closer to the Sun than when it is farther away, and should vary as
$1/L^2$. In the case of no neutrino oscillations or of the MSW solution
to the solar neutrino puzzle, the number of $^7$Be solar neutrino
induced events is supposed to vary according to the $1/L^2$ law,
following the variation of the total neutrino flux. 
This will be referred to as the ``normal'' seasonal
variation.

If vacuum oscillations are the solution to the solar
neutrino puzzle, large, anomalous
seasonal variations of the number of $^7$Be solar neutrino induced
events might be detected \cite{glashowkrauss,old_osc}. 
It is well known that neutrino oscillation effects depend on
the distance to the neutrino source, and 
different Earth-Sun distances may yield very different $\nu_e$
survival probabilities \cite{PP,sea_var}. 
The anomalous seasonal variation effect should 
be more pronounced in $^7$Be neutrinos than in $^8$B neutrinos (the
latter was recently studied in \cite{justsoback}). This
is due to one
important feature which distinguishes $^7$Be neutrinos 
from $^8$B and other abundant types of solar neutrinos: because they are
produced as part of a two-body final state, the neutrino energy
spectrum is mono-energetic.\footnote{In fact there are two distinct
  neutrino energies, 0.383 and 0.862 MeV, corresponding to different
  final states of the $^7$Li nucleus. Borexino and KamLAND are only
  sensitive to the higher energy component.} The details
will become clear when we discuss the anomalous seasonal variation
effect, in Sec.~\ref{sec:sensitivity}.

In the case of no anomalous seasonal variations,
if one has enough statistics and a small enough background, the
time variation of the data can be used to measure the solar
neutrino flux, given that the number of background events is
constant in time.\footnote{Actually, a time-dependent background is 
also acceptable, as long as it can be monitored and understood well 
enough.}  We will analyze how well
Borexino and KamLAND can perform this type of measurement. We are
particularly interested in analyzing the relevance of this technique
when the number of electron neutrinos reaching the detector is very
suppressed with respect to the Standard Solar Model predictions, as
might be
the case if there are $\nu_e\rightarrow \nu_{\mu,\tau}$ oscillations 
for the small angle MSW solution.\footnote{If $\nu_e$ oscillates into
  sterile neutrinos, the suppression is even more pronounced, due to
  the absence of neutral current $\nu_{\mu,\tau}$-$e$ elastic
  scattering. We do not consider oscillations into sterile neutrinos
  in this paper.}    

The paper is organized as follows. In Sec.~\ref{sec:flux} we discuss how
seasonal variations might be used to determine the solar neutrino flux
at Borexino and KamLAND,
in such a way that no separate measurement of the number of background
events is required. 
In Sec.~\ref{sec:sensitivity} we analyze the effect of the vacuum
oscillation solutions to 
the solar neutrino puzzle on the annual variation of the number of
detected events at Borexino and KamLAND. In particular we describe the
region of the ($\sin^22\theta,\Delta m^2$) parameter space where
vacuum oscillations can be discovered by studying the seasonal
variations of the data.
In Secs.~\ref{sec:measurement} and \ref{sec:exclusion} we describe how
the measurement of the seasonal 
variation of the $^7$Be solar neutrino flux may be used to either
measure the neutrino oscillation parameters, $\sin^22\theta$ and $\Delta
m^2$, or exclude a
large portion of the ($\sin^22\theta,\Delta m^2$) parameter
space. In Sec.~\ref{sec:conclusion} we summarize our results and conclude.

\setcounter{equation}{0}
\section{Measuring the $^7$Be Solar Neutrino Flux} 
\label{sec:flux}

As was already pointed out, measuring the flux of $^7$Be neutrinos is
crucial towards understanding the solar neutrino puzzle.
 Borexino \cite{Borexino}
plans to do this measurement by using 300 tons of organic liquid
scintillator to detect recoil electrons from elastic $\nu$-$e$
scattering.  Since the scintillator has no directional
information, and the signal is characterized only by the scintillation 
light produced by the recoil electron,
the background has to be kept under control.  This places a very 
stringent constraint
on the radio-purity of the scintillator and on the activity of
the material in the detector.  Borexino anticipates 100 tons of fiducial
volume for detecting solar neutrinos.  

KamLAND \cite{KamLAND}, which
was originally
conceived as a reactor neutrino experiment with an unprecedented
baseline (170~km on the average), may also be able to study $^7$Be solar neutrinos, 
if rigorous yet attainable requirements on the radio-purity and activity 
are met.  
We assume throughout the paper that
KamLAND will use 600 tons of fiducial volume for detecting solar
neutrinos (the size of the fiducial volume will depend on the
background rate, which is currently unknown). 
We concentrate our analysis on Borexino, which is an
approved dedicated solar neutrino experiment, and discuss KamLAND, whose
uses for solar neutrino studies are at present being proposed
\cite{kamlandprop}, 
as a possible higher statistics improvement. 

It is important to define what is meant by ``measuring the $^7$Be
solar neutrino flux.''
In reality, what the experiments are capable of measuring is 
the number of recoil electrons induced by solar neutrino
interactions in a given recoil electron kinetic energy range
(kinematic range). 
This information can only be converted into a solar
neutrino flux measurement if one knows the flavor composition of the
solar neutrinos \cite{dG_M}. Explicitly, assuming that the  
solar neutrino flux
is composed of $\nu_e$ (with fraction $P$) and $\nu_{\mu,\tau}$ (with
fraction $Q=1-P$), 
\begin{equation}
\#{\rm recoil}\hspace{1mm} 
{\rm electrons/time}
=\Phi\times(P\sigma_{\nu_e\mbox{-}e}+(1-P)\sigma_{\nu_{\mu,\tau}\mbox{-}e})N_e
,
\end{equation}
where $\Phi$ is the neutrino flux, $N_e$ is the number of target
electrons, and
\begin{equation}
\sigma_{\nu_{x}\mbox{-}e}\equiv\int_{T_{\rm
    min}}^{T_{\rm max}}dT \left(\frac{{\rm d}\sigma}{{\rm
    d}T}\right)_{\nu_{x}\mbox{-}e},
\end{equation}
with
$\left(\frac{{\rm d}\sigma}{{\rm d}T}\right)_{\nu_{x}\mbox{-}e}$ being
the
differential cross section for $\nu_x$-$e$ scattering for a given
kinetic energy $T$ of the recoil electron. $T_{\rm min}$ and $T_{\rm
  max}$  define
the kinematic range. 
In the case of neutrino oscillations, $P$ is the
the survival probability for electron neutrinos, while $1-P$ is the
probability that $\nu_{e}$ will oscillate into $\nu_{\mu,\tau}$. 

If the flavor composition of the flux is not
known, all that can be quoted is the effective neutrino flux,
$\Phi_{\rm eff}$, which is calculated from the number of measured recoil
electrons assuming that there are only electron neutrinos coming from
the Sun. Explicitly,
\begin{equation}
\Phi_{\rm eff}\equiv\frac{\#\rm{recoil}\hspace{1mm}
\rm{electrons/time}}{\sigma_{\nu_e\mbox{-}e}N_e}=\Phi\times\left( 
P+\left(1-P\right)\frac{\sigma_{\nu_{\mu,\tau}\mbox{-}e}}
{\sigma_{\nu_e\mbox{-}e}}\right).
\end{equation}
Clearly, if $P=1$, $\Phi_{\rm eff}=\Phi$. It is important to remember
that $\sigma_{\nu_{\mu,\tau}\mbox{-}e}/\sigma_{\nu_e\mbox{-}e}<1$ 
and therefore 
$\Phi_{\rm eff}\leq\Phi$. The ratio of the neutrino elastic cross
sections depends on the energy of the incoming neutrino and the
kinematic range to which each particular experiment is sensitive.
For $E_\nu=0.862$~MeV and the Borexino (KamLAND) kinematic range
250--800~keV (280--800~keV),  
$\sigma_{\nu_{\mu,\tau}\mbox{-}e}/\sigma_{\nu_e\mbox{-}e} = 0.213 (0.214)$. 
It is this effective electron
neutrino flux, $\Phi_{\rm eff}$, that is referred to, throughout this
paper (and in general), as the $^7$Be solar neutrino flux. 
 
In order to determine the number of recoil electrons induced by solar
neutrino interactions, it is crucial to determine
the number of background events. The number of background  
events can be estimated by various techniques, which we do not address
in this paper.  It is worthwhile to point out, however, that this is 
a very difficult process and it would be highly desirable to have an independent 
way to determine the $^7$Be solar neutrino flux in order to make the 
final results more convincing. This may be possible if one looks at
the seasonal variation of the number of detected events. 

In the following, we study the seasonal variation of the event rate as
a means to measure
the $^7$Be solar neutrino flux.  The distance between the Earth and
the Sun varies slightly over seasons because of the eccentricity of
the Earth's orbit.  The perihelion (when the Earth is closest to the
Sun) occurs around January first. 
The eccentricity of
the Earth's orbit is $\epsilon = 0.017$, and hence the distance varies
as
\begin{equation}
  L = L_0 ( 1 - \epsilon \cos (2 \pi t/{\rm year}) ),
\end{equation}
to the first order in $\epsilon$.  Here, $t$ is the time measured in years
from the perihelion, and $L_0 = 1.496 \times 10^8$~km is one
astronomical unit.  The neutrino flux varies as $1/L^2$ and hence
shows a seasonal variation of about 7\% from minimum to
maximum. The change in the Earth-Sun distance between the aphelion and
the perihelion is given by
\begin{equation}
\label{DeltaL}
\Delta L\equiv L_{{\rm max}}-L_{{\rm min}}=2\epsilon
L_0=5.1\times10^{6}\hspace{1mm}{\rm km}.
\end{equation}
By fitting the event rate to the seasonal variation expected
due to the eccentricity,
\begin{equation}
  B + S \left(\frac{L_0}{L}\right)^2,
\end{equation}
one can extract the background event rate $B$ and the signal event rate $S$
independently.  As long as the detector is monitored well and its
performance is sufficiently stable, this method will be only limited
by statistics.  

Borexino expects 53  
events/day\footnote{For simplicity, we neglect the contribution of solar
  neutrino sources other than $^7$Be electron capture throughout the paper.
  In particular we neglect the contribution of neutrinos produced
  in the CNO cycle, which is about 10\% of that from the $^7$Be
  neutrinos.} 
according to the BP95 \cite{BP95} Standard Solar Model (SSM),
together with 19 background events/day \cite{Borexino}, after the statistical
subtraction of the known background sources. 
This is done by pulse shape discrimination against the $\alpha$-particle
background and the measurement of Bi-Po pairs via $\alpha$-$\beta$
coincidence. This in turn allows the statistical subtraction of
processes in the
$^{238}$U and $^{232}$Th chains which are in equilibrium. 
It is also assumed that the experiment can achieve a radio-purity of
$10^{-16}$g/g for U/Th, $10^{-18}$g/g for $^{40}$K, $^{14}{\rm
  C}/^{12}{\rm C}= 10^{-18}$, and no Rn diffusion.  For KamLAND we
use 466 events/kt/day for the signal and 217 events/kt/day
\cite{kamlandprop} for the
background under similar assumptions but with larger cosmogenic
background (especially $^{11}$C) and some Rn diffusion.  
Assuming 600~t of fiducial volume, we expect 280 signal events/day and
130 background events/day.  
Throughout the paper, we will assume that the number of background
events is either constant
in time or its time dependence is sufficiently well understood by
monitoring. We neglect systematic effects and assume that there are
only statistical uncertainties. 

\begin{figure}[tp]
  \centerline{
    \psfig{file=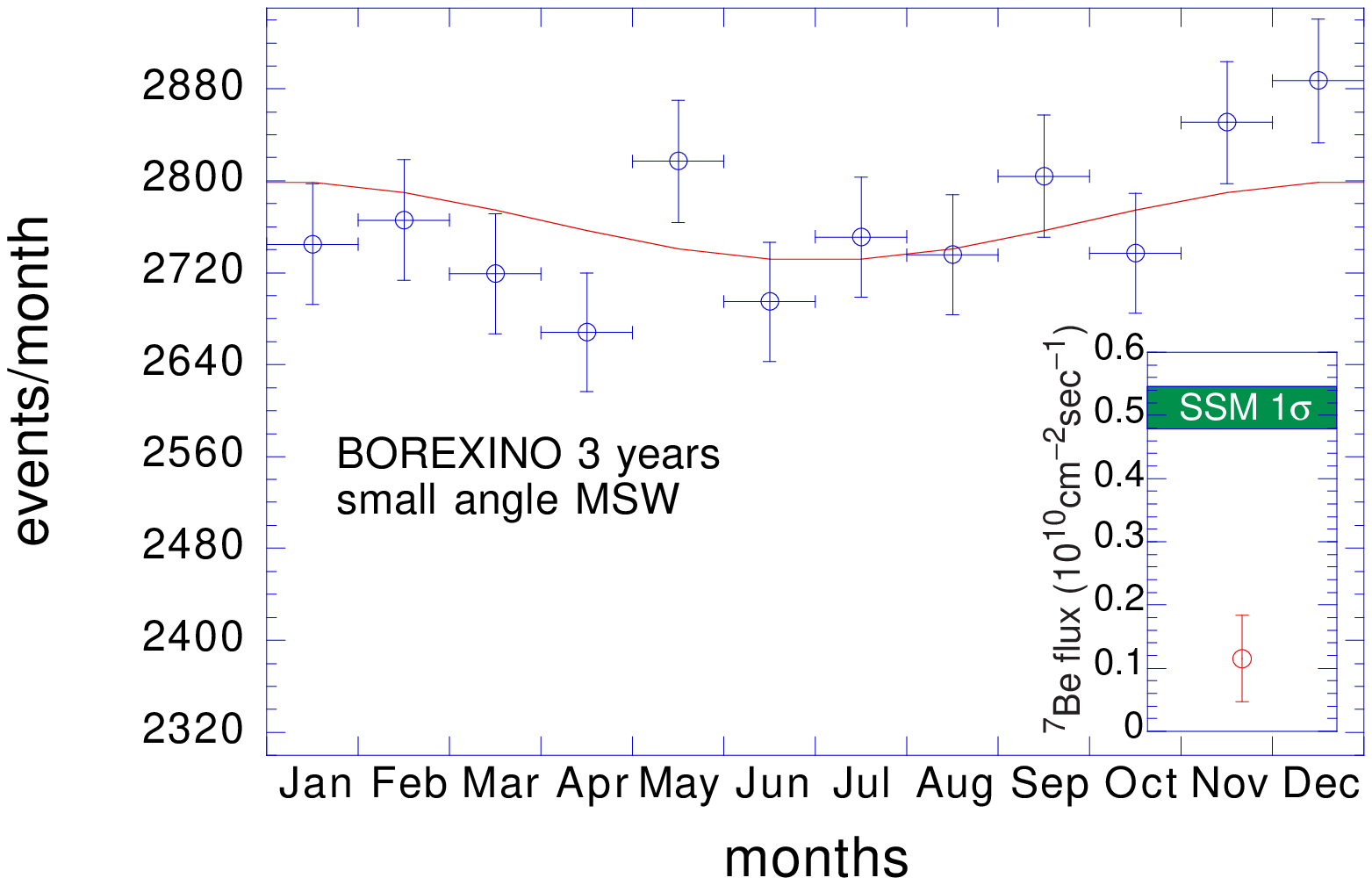,width=0.5\textwidth}
    \psfig{file=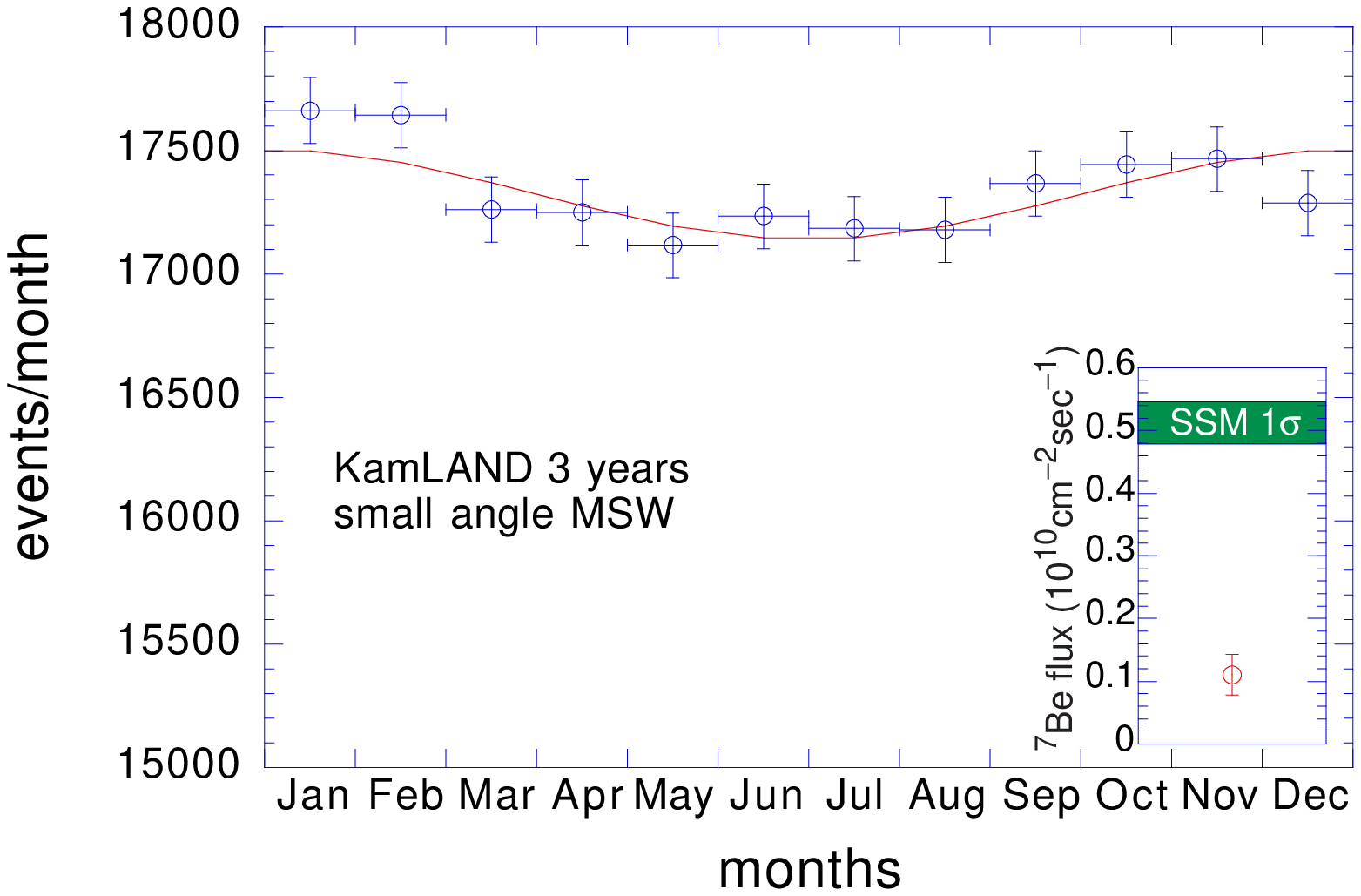,width=0.5\textwidth}
  }
\caption{The simulated seasonal variation of the $^7$Be flux for the case of the
    small angle MSW solution, for three years of Borexino (left) and
    KamLAND (right) running.  The inset shows the measured flux of
    $^7$Be neutrinos from the fit to the seasonal variation of the
    event rate (point with error bar) and the SSM prediction (shaded band).}
\label{seasonalMC} 
\end{figure}
Under these assumptions, Fig.~\ref{seasonalMC} depicts a simulation of
the seasonal variation of the ``data'' for both Borexino and KamLAND, after
three years of running.  The plots
are for the case of the small angle MSW solution to the solar neutrino
puzzle, where the $\nu_e$'s
produced by $^7$Be electron capture inside the Sun 
have almost completely oscillated into $\nu_\mu$ or $\nu_\tau$, 
and the event rate is reduced to 21.3\% (21.4\%) of the SSM prediction
at Borexino (KamLAND).  
In the fit to the data, both the
background and the $^7$Be flux are allowed to float. 

\begin{figure}[tp]
  \centerline{
    \psfig{file=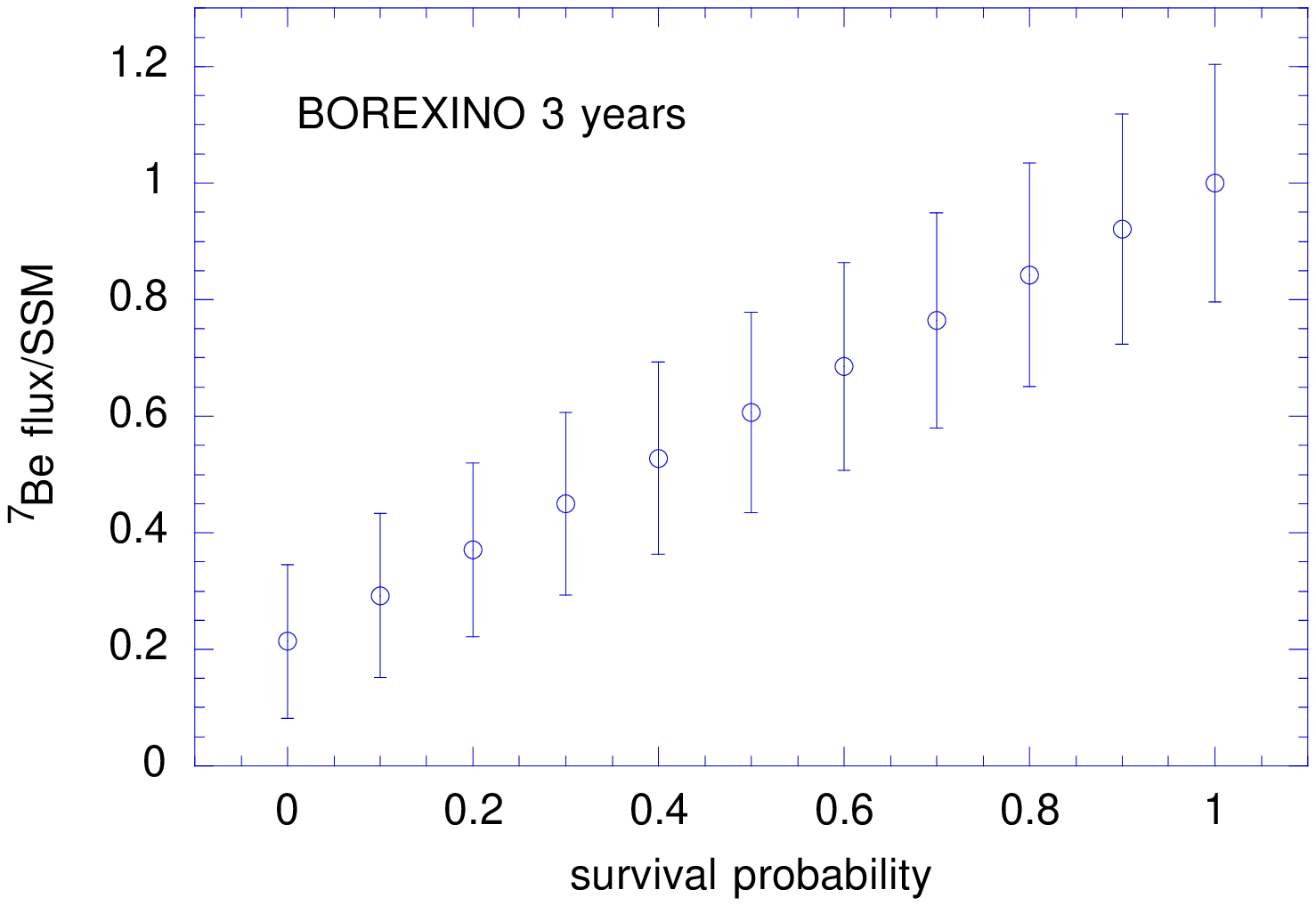,width=0.5\textwidth}
    \psfig{file=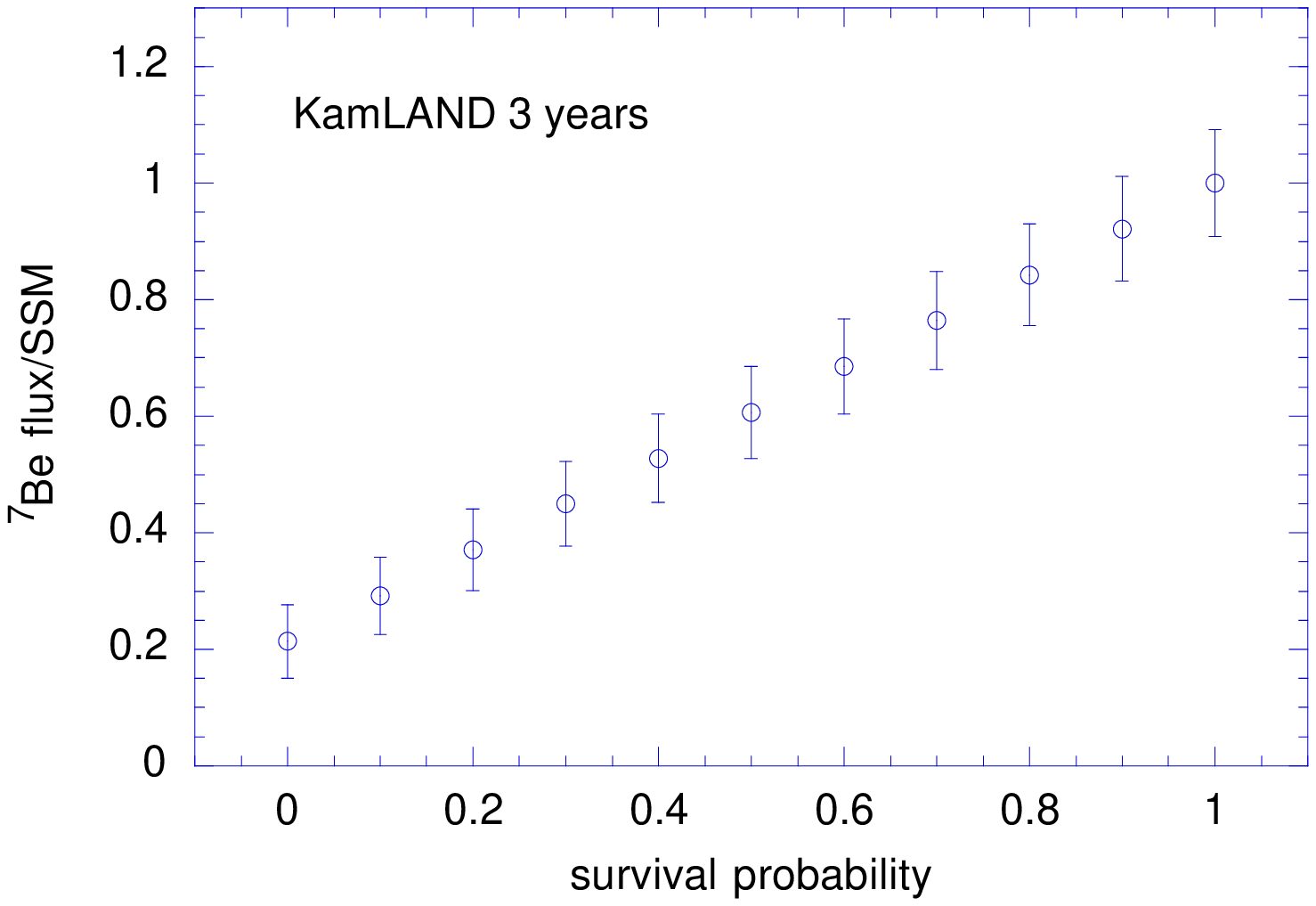,width=0.5\textwidth}
  }
\caption{The expected 1~$\sigma$ statistical accuracy of the $^7$Be neutrino flux
    measurement, together with the central value normalized by the flux 
         predicted by the SSM, as a function of the $\nu_e$ survival  
         probability at Borexino (left) and KamLAND (right), after
         three years of data taking.}
\label{flux-accuracy}
\end{figure}

This analysis can be repeated for different values of the $^7$Be flux, or,
equivalently, for different survival probabilities for
$\nu_e$. Fig.~\ref{flux-accuracy} depicts the expected 1~$\sigma$ statistical
accuracy of the $^7$Be flux measurement, together with the central value
normalized by the SSM prediction, as a function of the survival
probability for $\nu_e$. We emphasize that this measurement
technique assumes no knowledge of the background. 

The important information one should obtain from this analysis is if one can
indeed measure a nonzero $^7$Be solar neutrino flux.
For example, in the case of the small angle MSW solution, the $\nu_e$
survival probability is very close to zero and, assuming the expected
number of background events, Borexino's measured neutrino flux
is less than 1.5~$\sigma$ away from zero. The situation at KamLAND
is much better, and in the case of the small angle MSW solution a
healthy 3 sigma-away-from-zero measurement of the flux is obtained, if
the background is as low as expected. The significance of the measured
flux increases for larger survival probabilities, as in the case of
the large angle and the low $\Delta m^2$ MSW solutions.

\begin{figure}[tp]
  \centerline{
    \psfig{file=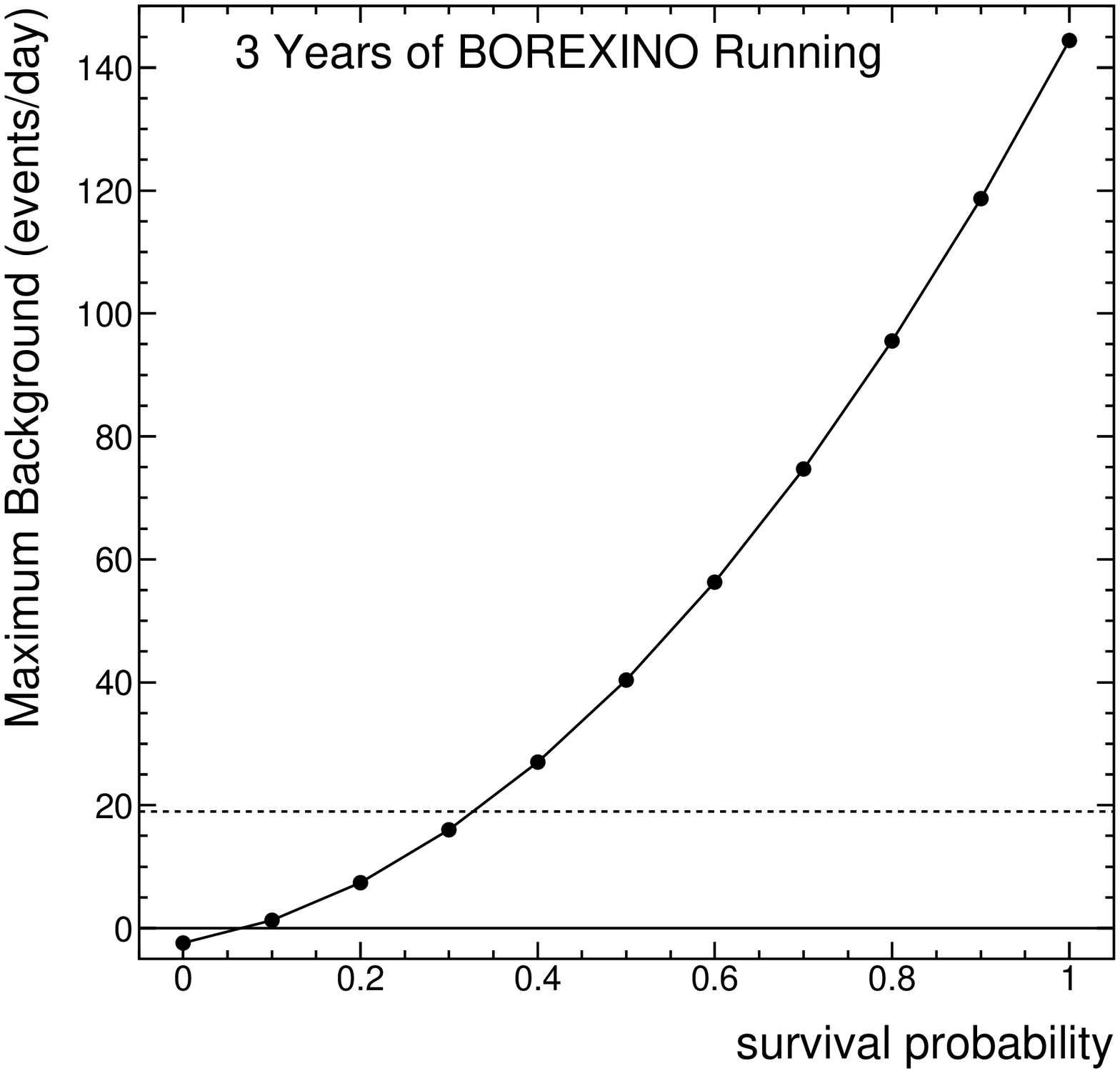,width=0.5\textwidth}
    \psfig{file=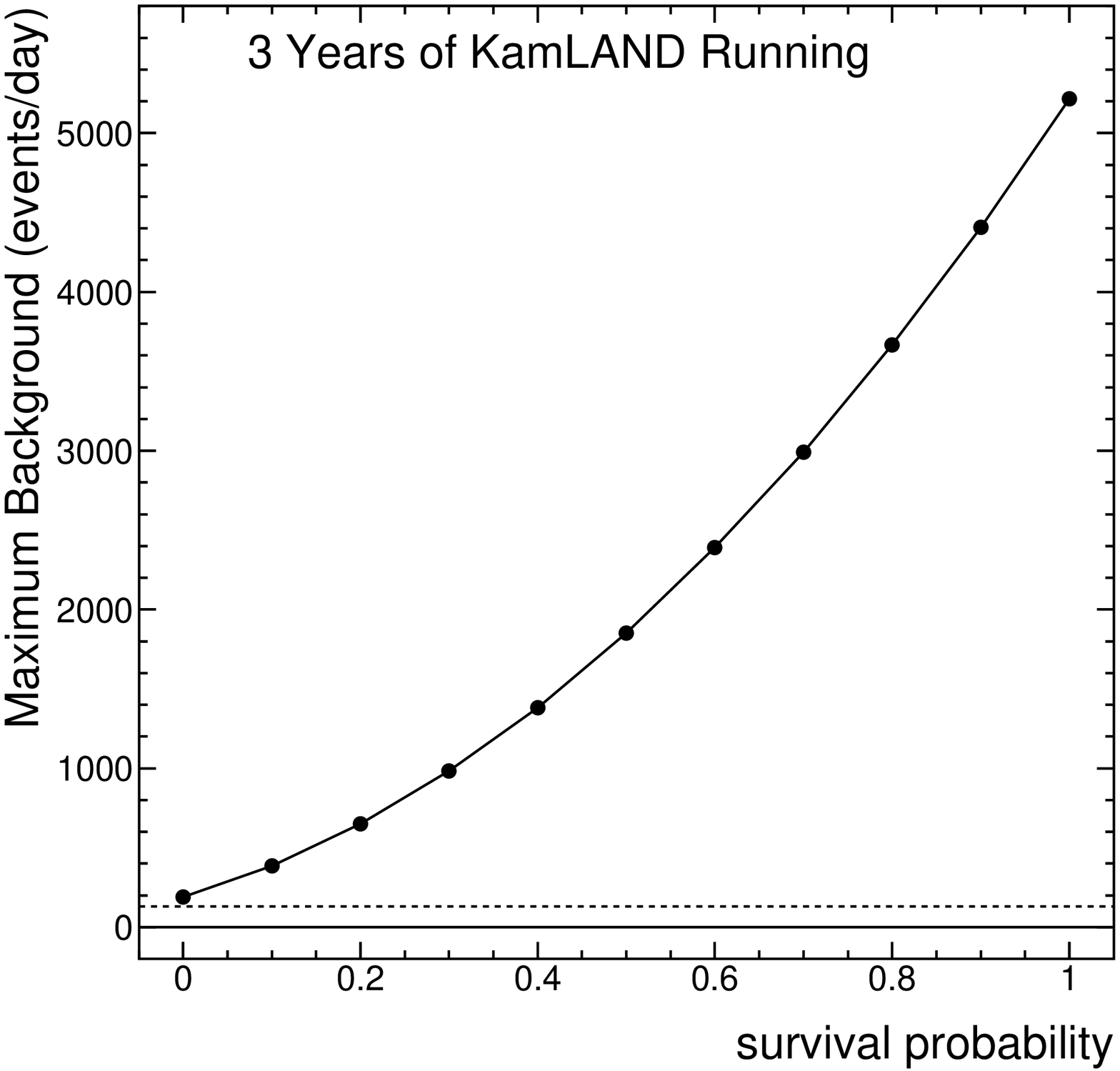,width=0.5\textwidth}
  }
\caption{The maximum number of background events allowed per day at Borexino
    (left) or KamLAND (right), for 3 years of running, in
    order to measure a solar neutrino flux which is 3~$\sigma$ away from
    zero. The dashed lines indicate the currently anticipated number of
    background events per day. }
\label{max_bkg}
\end{figure} 
A similar analysis can be performed in order to determine how many
background events each experiment can tolerate in order to claim
a solar neutrino flux measurement which is 3~$\sigma$ away from
zero. Fig.~\ref{max_bkg} depicts the maximum number of background
events per day allowed for 3 years of Borexino or KamLAND
running. It is worthwhile to comment that, in the case of Borexino and the
small angle MSW solution ($P\simeq0$), a 3 sigma-away-from-zero measurement of the
neutrino flux is not attainable
in three years, even in the case of no background (note that for
$P\lesssim0.05$ the required maximum background to achieve a three
$\sigma$ measurement of the flux is negative, {\it i.e.}, impossible to achieve). 
Therefore, for
Borexino, this simple, background independent analysis using the
seasonal variation of the data is not particularly powerful in the
case of the small angle MSW solution, 
due to statistical limitations. 

\setcounter{equation}{0}
\setcounter{footnote}{0} 
\section{Sensitivity to Vacuum Oscillations}
\label{sec:sensitivity}
 
In this section we study the discovery potential of the Borexino and 
KamLAND experiments in the region of $\Delta m^2$ corresponding to the 
vacuum oscillation solution to the solar neutrino problem. 
In this case, the pattern of seasonal variations can be very distinct 
from the normal pattern discussed in the previous section. 
 
The basic idea is the following. 
The survival probability $P$ for an electron neutrino in the case of
neutrino vacuum oscillations between two flavor states\footnote{One can
  assume the more complicated case of oscillations between three
  neutrino flavor states. In this paper we limit our studies to the
  case of oscillations between two flavor states.}  is given by 
\begin{equation} 
\label{osc} 
  P = 1 - \sin^2 2\theta \sin^2 \left(1.27 \Delta m^2 \frac{L}{E}\right), 
\end{equation} 
where the neutrino energy $E$ is in GeV, the distance $L$ in km, and the 
difference of masses-squared $\Delta m^2$ in eV$^2$. 
Model-independent analyses of all solar neutrino data show the 
need for an energy-dependent suppression of the $\nu_e$ flux. 
The ``just-so'' solution achieves this by choosing $\Delta 
m^2$ such that the corresponding neutrino oscillation length 
\begin{equation} 
  \label{eq:osclength} 
  L_{\rm osc}\equiv \frac{\pi E}{1.27\Delta m^2}=
  2.47 \times 10^8 \mbox{ km} \times 
  \left(\frac{E}{10 \mbox{ MeV}}\right) 
  \left(\frac{10^{-10} \mbox{ eV}^2}{\Delta m^2}\right) 
\end{equation} 
is of the order of one Astronomical Unit (1~a.u. = $1.496 \times 
10^8$ km); hence the name ``just-so''. More specifically, the 
oscillation length is assumed comparable to 1~a.u. for $^{8}$B 
neutrinos ($E_{\nu}\approx 10$~MeV); at the same time, the 
oscillation length of $^7$Be neutrinos ($E_{\nu} = 0.862$~MeV) is 
an order of magnitude smaller and, for sufficiently large $\Delta 
m^2$, can be comparable to the seasonal variation of the Earth--Sun 
distance due to the eccentricity of the Earth's orbit, $\Delta L$ (see
Eq.~(\ref{DeltaL})). As a 
consequence, the flux of $^7$Be neutrinos detected on the Earth may 
exhibit an anomalous seasonal variation, beyond the normal $1/L^2$ 
effect discussed in the previous section. 
 
Such anomalous variation could serve as a unique signature of 
vacuum oscillations \cite{glashowkrauss, old_osc}. 
Moreover, as we will show in this section, 
both Borexino and KamLAND will be able to cover a large portion of 
the ``just-so'' parameter space, even without relying on a 
particular solar model or estimate of the background rate, just by 
analyzing the {\it shape} of their data.
In this sense the discovery of an anomalous seasonal variation at one 
of these experiments would be as robust a result as the 
Super-Kamiokande measurement of the up-down asymmetry for the 
atmospheric muon neutrinos. 
 
\begin{figure}[t] 
 \centerline{ 
   \psfig{file=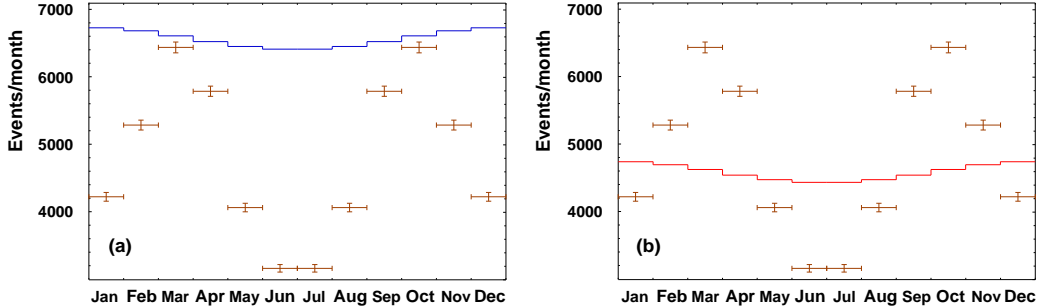,width=1\textwidth}} 
         \caption{ Illustration of the effect of 
           vacuum oscillations on the shape of the seasonal 
           variation of the solar neutrino data. 
           The points with statistical error bars represent the number of 
           events/month expected at Borexino after 3 years of running for 
           $\Delta m^2 = 3 \times 10^{-10}$~eV$^2$, $\sin ^2 2\theta = 
           1$. 
           The histogram in (a) shows the number of events predicted
           by the SSM without neutrino oscillations, plus the number of
           anticipated background events. 
           The histogram in (b) shows the same quantity 
           after adjusting the solar neutrino flux 
           and the background rate so as to 
           minimize the value of $\chi^2$, as explained in the 
           text. The difference between the case with oscillations
           and the one without oscillations is still apparent. } 
         \label{fig:illustratepoint} 
\end{figure} 
 
To illustrate the main idea, we choose a particular point 
($\Delta m^2 = 3 \times 10^{-10}$~eV$^2$, $\sin ^2 
2\theta = 1$) in the allowed region of the ``just-so'' parameter 
space\footnote{Based on the analysis of the total rates in the 
Homestake, GALLEX, SAGE, and Super-Kamiokande experiments. See 
Fig.~5 in \cite{bksreview}.} and compute the corresponding seasonal 
distribution of the neutrino events at Borexino after 3 years of 
running. We use the number of background events and the expected number of
signal events (before the effect of neutrino oscillations) quoted in
Sec.~\ref{sec:flux}.  
The results are shown in Fig.~\ref{fig:illustratepoint} by 
the set of ``data'' points with error bars; each point represents 
the number of events expected in a given month and the vertical error
bars show the corresponding statistical uncertainties. The histogram in 
Fig.~\ref{fig:illustratepoint}(a) shows ``theoretical'' event rates 
expected for non-oscillating neutrinos, provided the background rate
is known accurately and the SSM prediction for the neutrino flux is 
trusted. One can see that under these assumptions vacuum neutrino
oscillations with $\Delta m^2 = 3 \times 10^{-10}$~eV$^2$,  
$\sin ^2 2\theta = 1$ would be trivial to discover. 
 
More importantly, the experiment would be able to claim the 
discovery even without relying on an estimate of the background rate  
or the value of incoming neutrino flux predicted by the SSM. 
It is intuitively obvious from the figure that the 
vacuum oscillation ``data'' points cannot be fit by the 
``theoretical'' curve even if the background and the solar neutrino 
flux are varied freely, unless one assumes neutrino 
oscillations. This can be quantified as follows. For a given 
background rate $b$ and signal event rate $s$, we define the 
$\chi^2$ value of the fit for an ``average'' experiment: 
\begin{equation} 
\label{chi2} 
  \chi^2(s,b) = N_{\rm{d.o.f.}} + \sum_i^{N_{\rm{bins}}} \frac{(d_i - b - s\cdot h_i)^2}{d_i}, 
\end{equation} 
where $N_{\rm{bins}}$ is the number of bins, $N_{\rm{d.o.f.}}$ is the number of 
degrees of 
freedom, $d_i$ is the {\it average} expected number of neutrino 
events in the $i$th bin, and $h_i$ is given by $ h_i = \int_{i-1}^i ( 1 - 
\epsilon \cos (2 \pi x/N_{\rm{bins}}) )^2 dx$ . 
The constant term $N_{\rm{d.o.f.}}$ in Eq.~(\ref{chi2}) is added 
to take into account the effect of statistical fluctuations in the 
data. 
In a single experiment, statistical
fluctuations make the number of neutrino events in the $i$th
bin slightly different from $d_i$, and $\chi^2$ is computed by an
expression similar to Eq.~(\ref{chi2}), with $d_i$ replaced by the number of
events measured in the $i$th bin and {\it without} the constant term,
$N_{\rm d.o.f.}$.    
In our analysis, however, we are 
interested in the sensitivity of an ``average'' experiment. As 
proven in Appendix \ref{explainchi2}, averaging over many 
experiments results in the definition of $\chi^2$ given in
Eq.~(\ref{chi2}), with the constant term $N_{\rm{d.o.f.}}$. This agrees 
with the conventional wisdom that, if a function describes data 
correctly, the average expected value of $\chi^2$ should be equal 
to the number of degrees of freedom. Given this definition, we can 
choose values of $s$ and $b$ that minimize the $\chi^2$; the only 
restriction imposed is that both $s$ and $b$ be non-negative. For the case 
at hand the minimum occurs when $b$ is zero and $s$ is 
0.95 times the SSM prediction (see Fig.~\ref{fig:illustratepoint}(b)). 
As expected, even after this change  
the ``data'' points and the histogram are very different. 
(Numerically, $\chi^2=2935$ which for 10 degrees of freedom implies 
a confidence level of $1-9 \times 10^{-626}$!).\footnote{This number is,
  of course, unrealistic, and the true confidence level in this case will be
  dominated by systematic effects.} 
 
We now extend this approach, and scan the entire ($\sin ^2 2\theta$,  
$\Delta m^2$) plane 
(for an earlier work with a more simplified analysis which does
not consider the presence of background, see \cite{PP}).
In the analysis below, we follow the same steps 
as before: the ``data'' is simulated according to the expected number
of background and signal events, plus the effect of
neutrino oscillations, for each value of ($\sin ^2 2\theta$, $\Delta
m^2$), binned into a certain 
number of bins $N_{\rm{bins}}$, and then compared to the 
``theoretical'' predictions for the case of no oscillations. The 
$\chi^2$ is computed according to Eq.~(\ref{chi2}) and minimized with
respect to both the signal ($s$) and background ($b$). The confidence
level (CL) corresponding to the 
minimal value of $\chi^2$ and $N_{\rm d.o.f.}=N_{\rm{bins}}-2$ 
degrees of freedom is then determined, and the region in which the CL 
is less than a given number is isolated. This case, when both
the number of background events and the incoming solar neutrino flux
are considered unknown in the fit, is the most conservative one, and yields the
smallest sensitivity region.  Later we also study 
less conservative cases, where we assume in the ``data'' analysis that the
incoming neutrino flux is the one predicted by the SSM and/or that the
background rate is known. 
 
\begin{figure}[t] 
 \centerline{ 
   \psfig{file=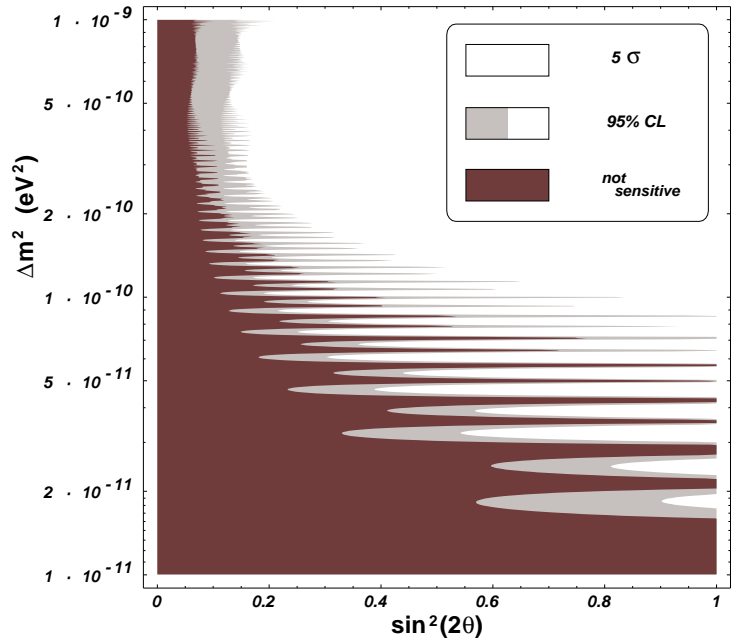,width=0.7\textwidth}} 
         \caption{The sensitivity region of the Borexino experiment 
           in 3 years, if the analysis does not assume any knowledge 
           of the background rate or the incoming solar neutrino flux. 
           In the unshaded region 
           the ``data'' is at least 5~$\sigma$ away from the best 
           no-oscillations fit. In the lightly shaded region the 
           discrepancy is greater than $95\%$~CL but less than 5~$\sigma$~CL.} 
         \label{fig:lowerBorex} 
\end{figure} 
 
We now apply this most conservative procedure to study the
experimental reach of Borexino after 3 years of operation. 
In Figure \ref{fig:lowerBorex} we show the results of the scan for 
$95\%$ and 5~$\sigma$ CL. As one can see from the 
figure, even at 5~$\sigma$ CL a large portion of the parameter space 
above $\Delta m^2 \sim 10^{-10}$~eV$^2$ is covered (white region). In this 
region the neutrino oscillation length $L_{\rm{osc}}$ is smaller than
the seasonal variation of the Earth--Sun distance $\Delta L$.
On the other hand, below $\Delta m^2 \sim 
10^{-10}$~eV$^2$ one can see a series of spikes protruding through the 
sensitivity region. It is important to understand the origin of these 
spikes. 
Since we adjust the level of signal and background in the fit, 
we are not sensitive to the absolute event rate, 
only to its variation during the year. For $\Delta m^2 \lesssim
10^{-10}$~eV$^2$ the oscillation length is larger than $\Delta L$ and the 
amplitude of the variation of the event rate is roughly proportional 
to the first derivative of Eq.~(\ref{osc}) with respect to $L$. In the 
regions where this derivative nearly vanishes, the amplitude of the 
variations is small and the signal is indistinguishable from the case 
of no oscillations. 
This explains why the loss of sensitivity occurs not only when the
neutrinos undergo approximately an integer number of oscillations as 
they travel to the Earth ($\Delta m^2=n\times 0.143\times
10^{-10}$~eV$^2$),
 but also when the number of oscillations is close to a
{\it half-integer} ($\Delta m^2=(n+1/2)\times 0.143\times
10^{-10}$~eV$^2$).
 In the latter case the absolute neutrino 
flux is maximally suppressed, but the magnitude of the seasonal 
variation is small.\footnote{Notice that the regions preferred from
the global fits have the absolute $^7$Be neutrino flux suppressed. See
Figs. \ref{fig:finalBorexino} and \ref{fig:finalKamLAND}.} 
 
Given this explanation, one would expect that the spikes corresponding 
to a  half-integer number of oscillations should become shorter if in the 
analysis we choose to rely on the SSM prediction of the incoming
neutrino flux and/or on the anticipated background rate.
It is straightforward to incorporate the knowledge of both
quantities and their uncertainties in our procedure. For
example, to impose the value of the incoming neutrino flux predicted
by the SSM, we 
modify the expression of $\chi^2$ in Eq.~(\ref{chi2}) by adding 
an extra term: 
\begin{equation} 
\label{modchi2} 
 \chi^2(s,b) \longrightarrow \chi^2(s,b) + \frac{(s-s_0)^2}{\sigma^2_{s_0}} , 
\end{equation} 
where $s$ and $b$ are the values of the signal and background with respect 
to which we later minimize $\chi^2$, $s_0$ is the SSM prediction for the 
signal, and $\sigma_{s_0}$ is the uncertainty in $s_0$. The rest of 
the analysis is carried out unchanged, except that the number of degrees of 
freedom is increased by one to $N_{\rm{d.o.f.}}=
N_{\rm{bins}} - 1$. To use both the incoming flux 
predicted by the SSM and the anticipated background rate, 
two terms are added to 
Eq.~(\ref{chi2}) and the number of degrees of freedom is increased 
by two to $N_{\rm{d.o.f.}}=N_{\rm{bins}}$. 

\begin{figure}[t] 
 \centerline{ 
   \psfig{file=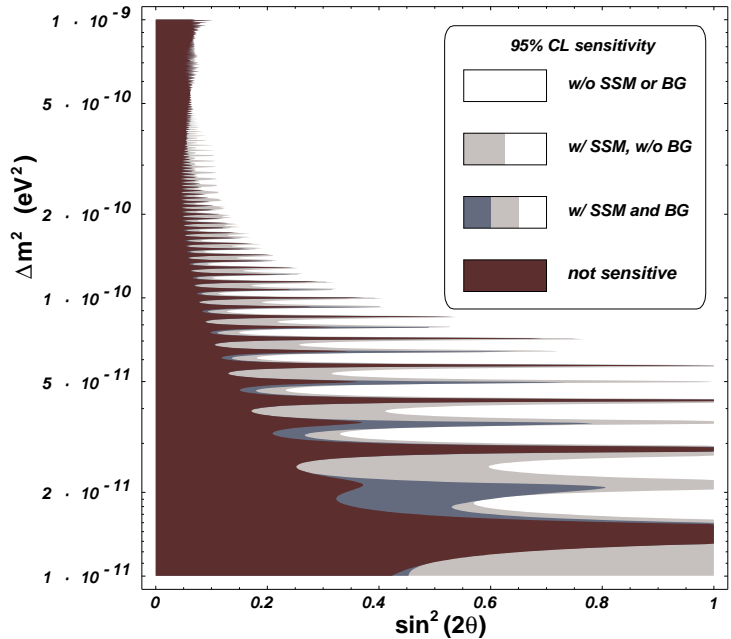,width=0.7\textwidth}} 
         \caption{The sensitivity reach of the Borexino experiment 
           after 3 years of running (at 95$\%$ confidence level). The 
           three cases considered are: no knowledge of either 
           the background rate or the incoming solar neutrino flux 
           (the covered region is white); assumption that the incoming
           solar neutrino flux is the one predicted by the SSM, with 
           9$\%$ uncertainty (the covered 
           region is white + light gray); assumption that the
           background rate is known with 10\% uncertainty and the
           incoming neutrino flux agrees with the SSM, with 9\%
           uncertainty (the covered region is white + light 
           gray + medium gray).} 
         \label{fig:lowerBorexssmbg} 
\end{figure}

The results of the calculation are shown in Fig.~\ref{fig:lowerBorexssmbg}. 
The uncertainty on the solar model prediction of the 
$^7$Be neutrino flux is taken to be $9\% $ \cite{bp98}, while the 
uncertainty on the background is $10\%$ \cite{kamlandprop}.
As expected, the odd--numbered spikes do become shorter. The one
possibility not shown in the plot is the situation when one only assumes
knowledge of the background rate. In this case the spikes become
significantly thinner, although their length remains virtually unchanged.
 
\begin{figure}[t] 
 \centerline{ 
   \psfig{file=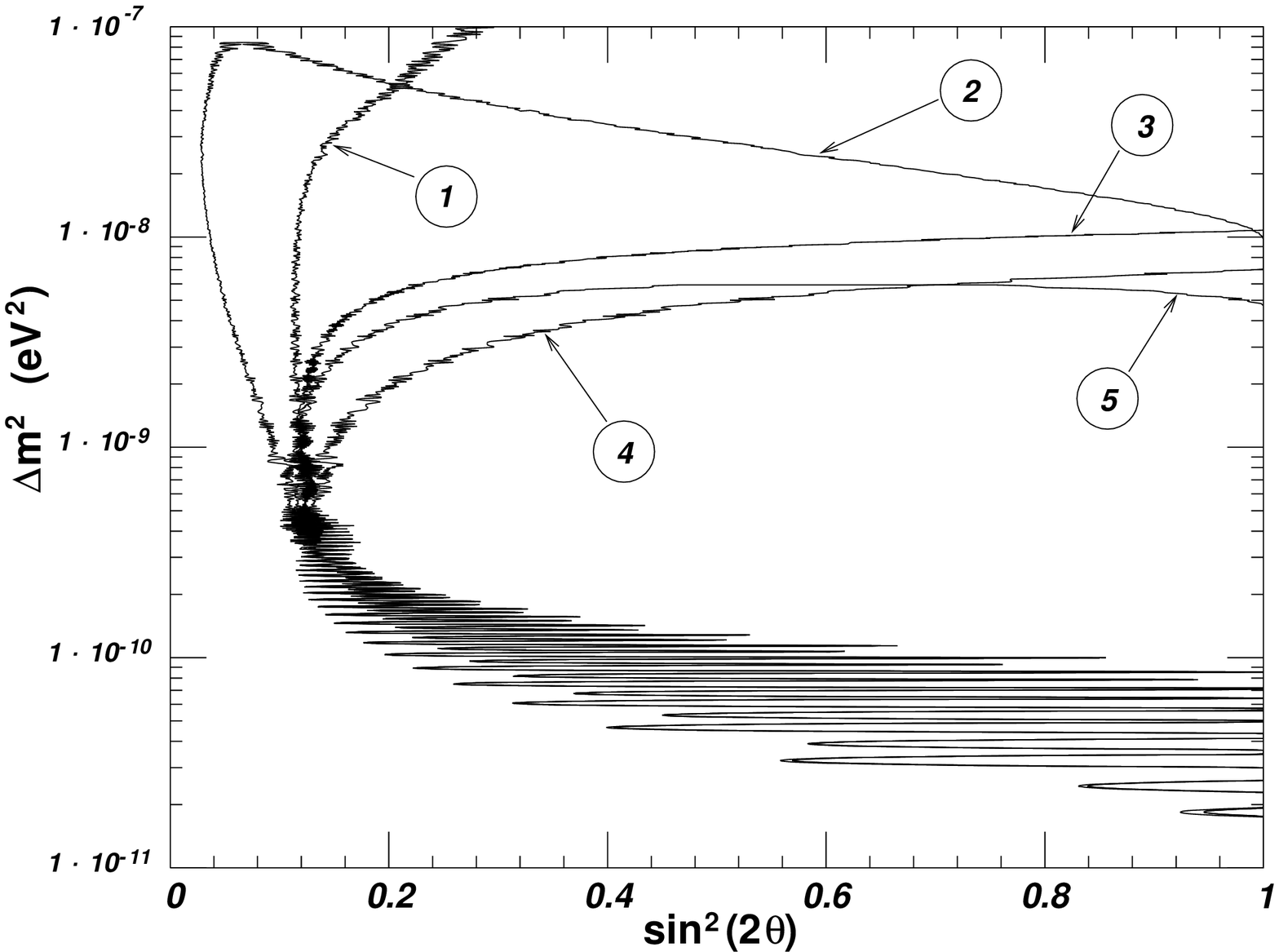,width=0.7\textwidth}} 
         \caption{The relative roles of the binning effect, the 
           linewidth effect, and the matter effect, as 
           explained in the text. } 
         \label{fig:lnwdthandcore} 
 \end{figure}

In order to extend this analysis to values of $\Delta 
m^2 > 10^{-9}$~eV$^2$, several issues must be confronted. 
We will next address these issues one by one, and illustrate 
the discussion in Fig.~\ref{fig:lnwdthandcore}. 
 
The first and the most obvious point is that the number of bins needs to be 
changed. The reason is that the frequency of the seasonal variations
increases with $\Delta m^2$, and above some value ($\Delta
m^2\simeq 8 \times 
10^{-10}$~eV$^2$, for 12 bins) integration over the bin size washes out 
the effect. To avoid this, we change the number of bins from 12 to 
365. After the change, the effect of binning 
kicks in at $\Delta m^2\simeq 2.4 \times 10^{-8}$~eV$^2$, as curve 1 in  
Fig.~\ref{fig:lnwdthandcore} illustrates. 
 
Next, there are two physical effects one must take 
into account: one is the interaction of the neutrinos with solar
matter (the MSW effect), and the other is
the finite width of the $^7$Be solar neutrino line. One may worry
about the wash-out of the seasonal variation effect due to the finite
size of Sun's core. However, matter effects 
make the core size effect irrelevant because
the mixing angle in the Sun's core is small and the oscillations
effectively start at the level-crossing point (see 
Eq.~(\ref{eq:arrival})).\footnote{We thank E.~Lisi and L.~Wolfenstein 
for pointing this out to us. For earlier papers on this particular
point, see \cite{W}, \cite{glashowkrauss}, and in particular,
\cite{Pantaleone}.} 

\begin{figure}
  \centerline{
  \psfig{file=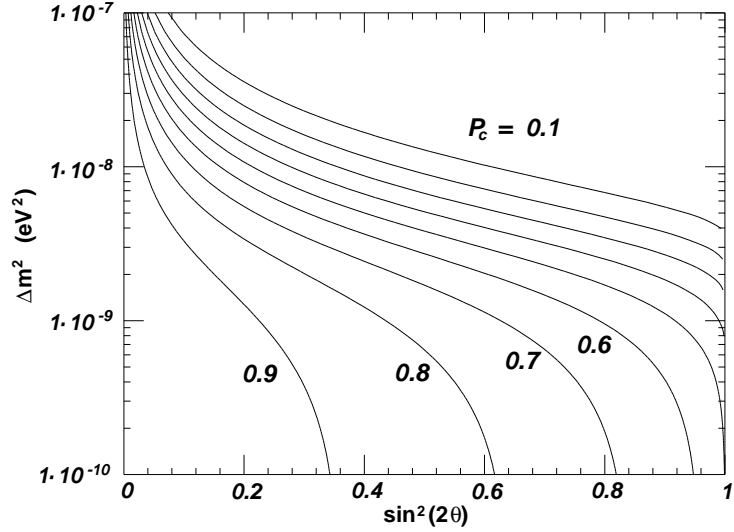,angle=90,width=0.7\textwidth}}
  \caption{The contour plot of the hopping probability $P_c = 0.1$,
    0.2, $\ldots$, 0.9, for the $^7$Be neutrino energy, using the
    exponential-profile approximation for the electron number density
    and Eq.~(\ref{eq:Petcov}).}
  \label{fig:Pc}
\end{figure}

\begin{figure}[tp] 
 \centerline{ 
   \psfig{file=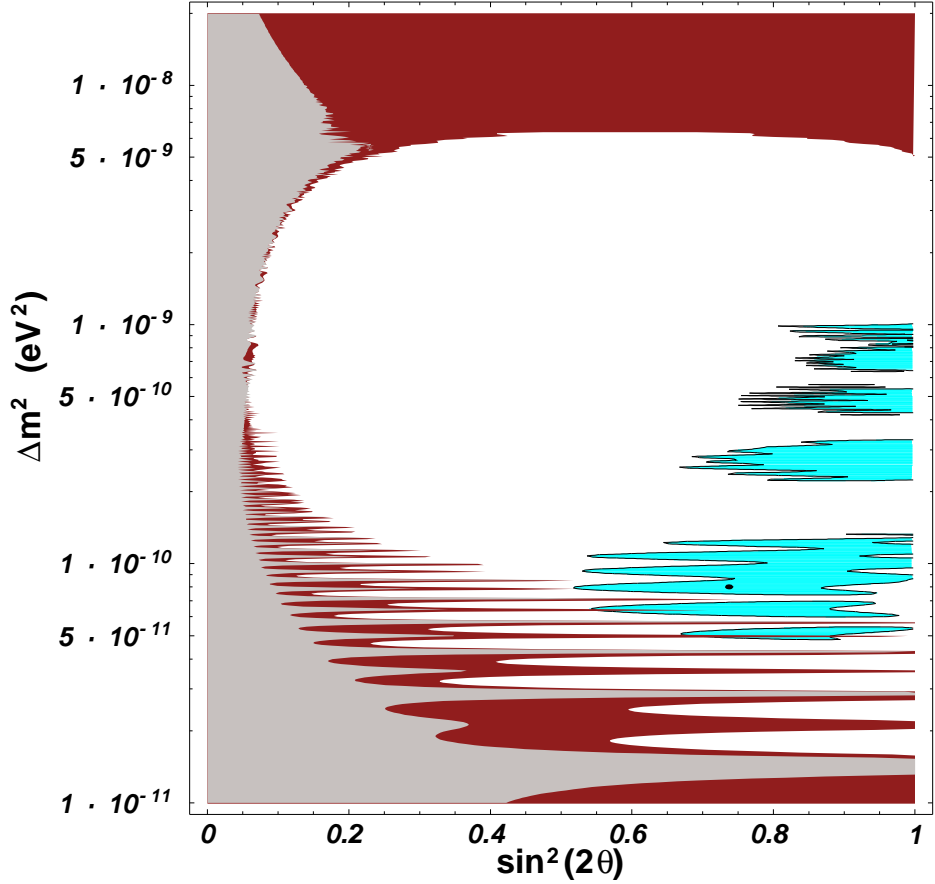,width=0.9\textwidth}} 
         \caption{The final sensitivity plot for three years of
           Borexino running, after 
           the inclusion of all effects limiting the reach of the 
           experiment for large $\Delta m^2$. The white region corresponds 
           to the sensitivity at more than 95$\%$ confidence level
           with both the incoming neutrino flux and background rate
           assumed to be unknown, and the dark region to the
           additional coverage when the SSM $^7$Be flux and the
           background rate estimated elsewhere are used. Also shown are 
           the regions preferred by the analysis of the total 
           rates in the Homestake, GALLEX, SAGE, and Super-Kamiokande 
           experiments \cite{bksreview}.} 
         \label{fig:finalBorexino} 
 \end{figure}

\begin{figure}[tp] 
 \centerline{ 
   \psfig{file=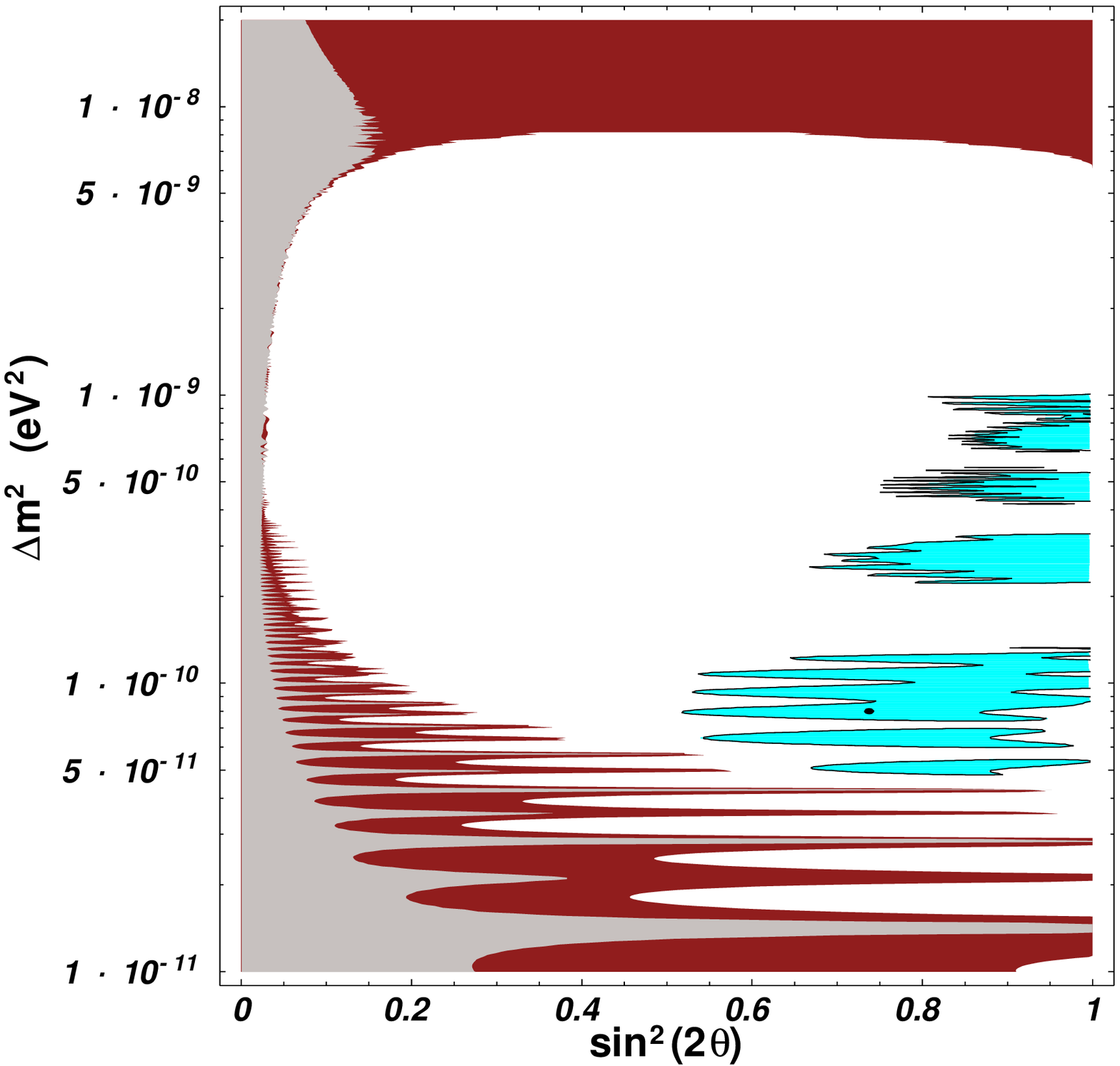,width=0.9\textwidth}} 
         \caption{The same as Fig \ref{fig:finalBorexino}, but for
           three years of KamLAND running.} 
         \label{fig:finalKamLAND} 
 \end{figure}

When a $\nu_e$ is created by the electron capture process
in the core of the Sun, its Hamiltonian is dominated by the matter
effect $\sqrt{2} G_F n_e$ ($n_e$ is the electron number density) if
$\Delta m^2 \ll 10^{-5}~{\rm eV}^2$ for $^7$Be neutrinos.
We restrict ourselves to $\Delta m^2 < 10^{-7}$~eV$^2$ in the
following discussions, as the final sensitivity due to the anomalous
seasonal variation is limited by $\lesssim
10^{-8}$~eV$^2$ as will be seen later in this section.  Then the mass
mixing  effect can
be completely ignored at the time of the neutrino production, and one
can safely take the produced neutrino to be in a Hamiltonian
eigenstate (the one which corresponds to the larger energy in the
Sun's core).  
As it propagates through the Sun, the neutrino follows the instantaneous
Hamiltonian eigenstate (in the adiabatic approximation), and exits in
the heavier mass eigenstate, $\nu_2 = \nu_e \sin \theta + \nu_\mu \cos
\theta$. It also has a finite amplitude $A_c$ for hopping to the other
Hamiltonian eigenstate.  
The neutrino state that exits the Sun can therefore be written as
\begin{equation}
  \nu_{\rm exit} = A_c \nu_1 + B_c \nu_2,
\end{equation}
with the unitarity constraint $|A_c|^2 + |B_c|^2 = 1$.  Out of the
Sun, the two mass eigenstates develop different phases due to the
mass difference, $e^{-i \Delta m^2 t / 2 E_\nu}$.  Therefore the
neutrino state that arrives at the Earth is given by
\begin{equation}
  \nu_{\rm arrival} = A_c \nu_1 + B_c \nu_2 e^{-i \Delta m^2 L/2E_\nu},
  \label{eq:arrival}
\end{equation}
up to an overall phase factor.
The distance $L$ is between the point of level crossing and the
Earth.  Finally, the survival probability of the electron neutrino is
determined by the $\nu_e$ component of $\nu_{\rm arrival}$, and
hence
\begin{eqnarray}
  P &=& |A_c \cos \theta + B_c \sin \theta e^{-i \Delta m^2 L/2E_\nu}|^2
  \nonumber \\
  & = & |A_c|^2 \cos^2 \theta + |B_c|^2 \sin^2 \theta
  + 2 {\rm Re} A_c^* B_c e^{-i \Delta m^2 L/2E_\nu} \sin\theta \cos\theta .
\end{eqnarray}
Since $|B_c|^2$ is the hopping probability between two Hamiltonian
eigenstates in the Sun $P_c$, one can rewrite the formula using $P_c$
and an additional phase factor $A_c^* B_c = \sqrt{P_c (1-P_c)}
e^{-i\delta}$,
\begin{equation}
  P = P_c \cos^2 \theta + (1-P_c) \sin^2 \theta
  + 2 \sqrt{P_c (1-P_c)} \sin\theta \cos \theta
  \cos \left( \frac{\Delta m^2 L}{2E_\nu} + \delta \right).
    \label{eq:withMSW}
\end{equation}
An approximate formula for $P_c$ was given in \cite{Petcov} using the
exponential density profile of the Sun,
\begin{equation}
  P_c = \frac{e^{-\gamma \sin^2 \theta}-e^{-\gamma}}{1-e^{-\gamma}}
  \label{eq:Petcov}
\end{equation}
with
\begin{equation}
\gamma = 2\pi r_0\frac{\Delta m^2}{2 E_{\nu}}=1.22\left(\frac{\Delta
  m^2}{10^{-9}{\rm eV}^2}\right)\left(\frac{0.862{\rm MeV}}{E_{\nu}}\right),
\end{equation}
where we consider the exponential-profile approximation for the 
electron number density in the Sun $n_e \propto \exp(-r/r_{0})$, with
$r_0=R_\odot/10.54=6.60\times 10^{4}$~km, given in \cite{Bahcallbook}.
Fig.~\ref{fig:Pc} shows the contours of $P_c$ on the $(\sin^2 2\theta,
\Delta m^2)$ plane for the $^7$Be neutrino energy $E_\nu = 0.862$~MeV.

The most important consequence of the matter effect is that the vacuum
oscillation is suppressed when $P_c \rightarrow 0$ (adiabatic limit).
The origin of the suppression is simple.  When $P_c$ is small, the
neutrino state that exits the Sun is nearly a pure $\nu_2$ state.  Since
it is a mass eigenstate, only its phase evolves in time and no
oscillations take place.  The $\nu_e$ survival probability then is
simply given by the $\nu_e$ content of $\nu_2$, which is nothing but
$\sin^2 \theta$, without anomalous seasonal variations.  Therefore, the
sensitivity to the anomalous seasonal variation is reduced in the region
with small $P_c$.  When $\Delta m^2$ is small, on the other hand, the situation
is in the extreme non-adiabatic limit, and $P_c \rightarrow \cos^2
\theta$.  Then Eq.~(\ref{eq:withMSW}) reduces
to Eq.~(\ref{osc}).  As $\Delta m^2$ increases, $P_c$ becomes smaller
than $\cos^2 \theta$, which enhances the vacuum oscillation effect in
the small mixing angle region.  Curve 2 in
Fig.~\ref{fig:lnwdthandcore} includes the matter effect and indeed
indicates a reduced sensitivity for large $\sin^2 2\theta$ (small $P_c$)
and an enhanced sensitivity for small $\sin^2 2\theta$ (where $P_c$
starts deviating from $\cos^2 \theta$).\footnote{In the numerical
  scan, we ignored the additional phase factor $\delta$, because its
  effects are negligible \cite{Pantaleone}.}

The second effect is the finite width of the $^7$Be line. To give some 
preliminary 
idea about the relative size of this effect, we first consider a simplified 
model. We assume for a moment that the only source of the line 
broadening is the Doppler shift of 
neutrino energies arising from the thermal motion of the $^7$Be 
nuclei. Since the energy is shifted to $E\rightarrow E(1+v_z/c)$
and the probability distribution of the velocity along the line of
sight $v_z$ is proportional to $\exp(-m
v_z^2/2kT)$, the resulting line profile will be a Gaussian $\exp(-m
c^2 (E-E_0)^2/(2kT E_0^2))$. 
Taking the temperature to be 15.6 million Kelvin (the temperature 
in the center of the Sun) and integrating over the
line profile, we obtain curve 3 in Fig.~\ref{fig:lnwdthandcore}. The sensitivity loss 
now occurs at $\Delta m^2\simeq 1 \times 10^{-8}$~eV$^2$,
demonstrating that this effect is more 
important than the matter effect.
 
This naive model is actually incomplete; there exists another very 
important source of line broadening. Because the incoming electron in 
the process $^7\mbox{Be} + e^- \rightarrow {^7\mbox{Li}} + \nu_e$ 
has nonzero thermal kinetic energy, the center of mass energy of 
the reaction is greater than the one measured in the laboratory, and so the 
neutrino has a greater energy. 
The phase space distribution of electrons is governed by the
Maxwellian factor $\exp(-E_{e^-}/k T)$. This distribution has to be
multiplied by the energy-dependent cross section, integrated over the
phase space, and finally convoluted with the Gaussian arising from the
Doppler effect.
The resulting line 
shape becomes asymmetric, with a Gaussian profile on the left (due to 
the Doppler effect) and an exponential tail on the right (due to the 
Maxwellian distribution of the electron energy). The issue was studied 
in detail in \cite{lineprfl}, where the precise form of the profile was 
computed.\footnote{It turns out that other effects, such as 
collisional line broadening \cite{loeb} or gravitational
energy shift \cite{lineprfl}, are unimportant.} 
Repeating the calculation with this profile we generate curve 4 in  
Fig.~\ref{fig:lnwdthandcore}. 
 
One can see that for this curve the cut-off 
occurs at smaller $\Delta m^2$. This behavior 
is expected, because the linewidth is now greater than when only the 
Doppler effect was included (curve 3 in
Fig.~\ref{fig:lnwdthandcore}). 
It is also worth noting that the cut-off 
sets in more gradually. This feature can be understood 
analytically by considering the Fourier transform of an exponential tail 
vs. a Gaussian tail. The details can be found in Appendix \ref{app2}. 
 
Finally, we can combine both the linewidth and the matter
effects. The result is curve 5 in Fig.~\ref{fig:lnwdthandcore}.    
As expected, the inclusion of the matter effect on top of the
linewidth effect introduces only a small distortion to the sensitivity
region.  It is important to
note that for $\Delta m^2\lesssim 5\times 10^{-10}$~eV$^2$ none of the
physical effects mentioned above affect the sensitivity region (curve
1 versus curve 5, in Fig.~\ref{fig:lnwdthandcore}).  
 
We need to consider one last ingredient in the analysis. We 
again return to the issue of the number of bins. While choosing more 
bins is necessary for larger values of $\Delta m^2$, it simultaneously 
leads to a loss of sensitivity for smaller $\Delta m^2$. 
A better procedure is to use an optimum number of bins 
$N_{\it opt}$ for each $\Delta m^2$. It can be shown that for our method of 
analysis (minimizing $\chi^2$ by varying the signal and background) 
and sufficiently large $\Delta m^2$ an 
approximate formula holds: $N_{\it opt} \simeq 2 \times 10^{10} (\Delta m^2 
/1\mbox{ eV}^2)$.  Of course, this formula should not be used when 
the optimal number of bins it predicts is too small. 
We choose to use 12 bins for 
$\Delta m^2 \leq 6 \times 10^{-10}$~eV$^2$ and a variable
number of bins $N_{\rm{bins}} = 2 \times 10^{10} (\Delta m^2 
/1\mbox{ eV}^2)$ for $\Delta m^2 > 6 \times
10^{-10}$~eV$^2$.\footnote{An alternative technique, which can be
considered more rigorous but 
which would also be more computer intensive, is to Fourier transform 
the simulated data for every value of ($\sin^22\theta,\Delta m^2$) in
the scan. One can then compare the intensities of the harmonics to
those expected for the case of no oscillations. A 
description of this method can be found in \cite{fourier}. For our 
purposes varying the number of bins is sufficient.}
 
In Fig.~\ref{fig:finalBorexino} we show the entire sensitivity reach
of Borexino after three years of running. 
The unshaded region will be covered at
least at 95$\%$ CL, if in the analysis one allows the background and
the incoming solar neutrino flux to float. The 
dark shading marks the additional portion of the parameter space that
will be covered at least at 95$\%$ CL, if in the analysis one assumes both the 
anticipated background rate (10$\%$ uncertainty) and the SSM
prediction of the $^7$Be solar neutrino flux (9$\%$ uncertainty).  
For $\Delta m^2 \gtrsim 5 \times 10^{-9}~{\rm eV}^2$, the sensitivity
to the anomalous seasonal variation gets lost because of the
smearing due to the linewidth effect.  However, there is an overall
suppression of the flux due to the MSW effect in this region.
To be sensitive to this overall suppression, we should return
to a smaller number of bins to enhance the statistical accuracy.  We
therefore use 12 bins in this region.\footnote{One can cover a
  slightly larger portion of the parameter space by using yet fewer
  bins.  We chose 12 bins such that one can still verify the expected 
  $1/L^2$ behavior of the signal even with a reduced flux, as we
  discussed in Sec.~\ref{sec:flux}.}

For comparison, we also superimpose 
the ``just-so'' preferred regions obtained by analyzing the total event
rates in the Homestake, GALLEX, SAGE, and Super-Kamiokande  
experiments (Fig.~5 in \cite{bksreview}). The plot shows 
that Borexino will be sensitive to almost all of the preferred
region, even without relying on the SSM prediction of the incoming
neutrino flux or on the knowledge of the background rate. Only two thin spikes 
protrude through the lower ``islands''. This overlap disappears 
completely when the anticipated background rate and the SSM prediction
for the incoming neutrino flux are used in the ``data'' analysis, 
in which case the entire preferred region is covered. 

Fig.~\ref{fig:finalKamLAND} contains a similar plot for three years of 
KamLAND running. Because KamLAND will have more statistics, it will be
sensitive at 
$95\%$ CL to the entire preferred region without relying in the
analysis on the SSM prediction of the incoming neutrino flux or on  
the knowledge of the background rate. 

As mentioned earlier, the
sensitivity to anomalous seasonal variations is completely lost for
$\Delta m^2\gtrsim 10^{-8}$~eV$^2$. In this case 
the seasonal variation of the data is consistent with an average
suppression of the incoming neutrino flux. In particular, in the case
of the MSW solutions ($10^{-7}\mbox{ eV}^2\lesssim\Delta m^2\lesssim
10^{-4}\mbox{ eV}^2$), no anomalous seasonal variations can be
detected, as was implicitly assumed in Sec.~\ref{sec:flux}.

At last, it is worth mentioning that the experiments will still be
sensitive to a significant part of the preferred region even if the 
background rate or the incoming $^7$Be neutrino flux (for all flavors)
turns out to be significantly different.  
For example, if the background rate at Borexino (KamLAND) turns out to be
30 (100) times higher than expected, the part of the preferred region with 
$\Delta m^2 > 10^{-10}$~eV$^2$ will still be within the reach of the
experiment, after three years of running. 
The sensitivity will be completely lost only if the background rate turns
out to be three (four) orders of magnitude higher than anticipated at
Borexino (KamLAND). 
The consequences of a $^7$Be solar
neutrino flux smaller than predicted by the SSM can also be studied.  
If the $^7$Be neutrino flux is for some reason
suppressed by a factor of 5, KamLAND is still sensitive to the part of
the preferred region with $\Delta m^2 > 10^{-10}$~eV$^2$, after 3
years of running.
 
\setcounter{equation}{0}
\section{Measuring the Oscillation Parameters}
\label{sec:measurement}

\begin{figure}[t]
 \centerline{
   \psfig{file=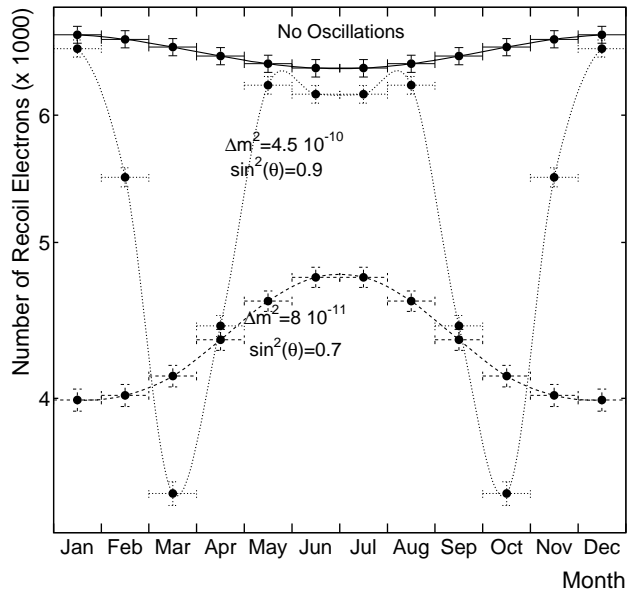,width=0.6\textwidth}
 }
         \caption{Number of recoil electrons detected in a given
   month, for the low point, the high point (see text for description)
   and the case of no neutrino oscillations, after three years of
   Borexino running.}
\label{points}
\end{figure}

In this section, we address the issue of how well the two-neutrino
oscillation parameters, $\sin^22\theta$ and $\Delta m^2$, can be
extracted if the data collected at future solar neutrino experiments
exhibits an anomalous seasonal variation. 
In order to do this, we simulate ``data'', according to the procedure
developed in Sec.~\ref{sec:sensitivity}, for two distinct points in
the parameter 
space, $\sin^22\theta=0.7$, $\Delta m^2=8\times
10^{-11}$~eV$^2$ (``low point'') and $\sin^22\theta=0.9$,
$\Delta m^2=4.5\times 10^{-10}$~eV$^2$ (``high point''). 
The low point is close to the best
fit point presented in \cite{bksreview}, while the high point is close
to the point preferred by the Super-Kamiokande analysis of the recoil 
electron energy
spectrum \cite{Super-K_spectrum}. The data is binned into months (12
bins per years), and Fig.~\ref{points} depicts the annual variations
for both the high and the low points, assuming three years of Borexino
running. The no-oscillation case is also
shown.

\begin{figure}[t]
 \centerline{
   \psfig{file=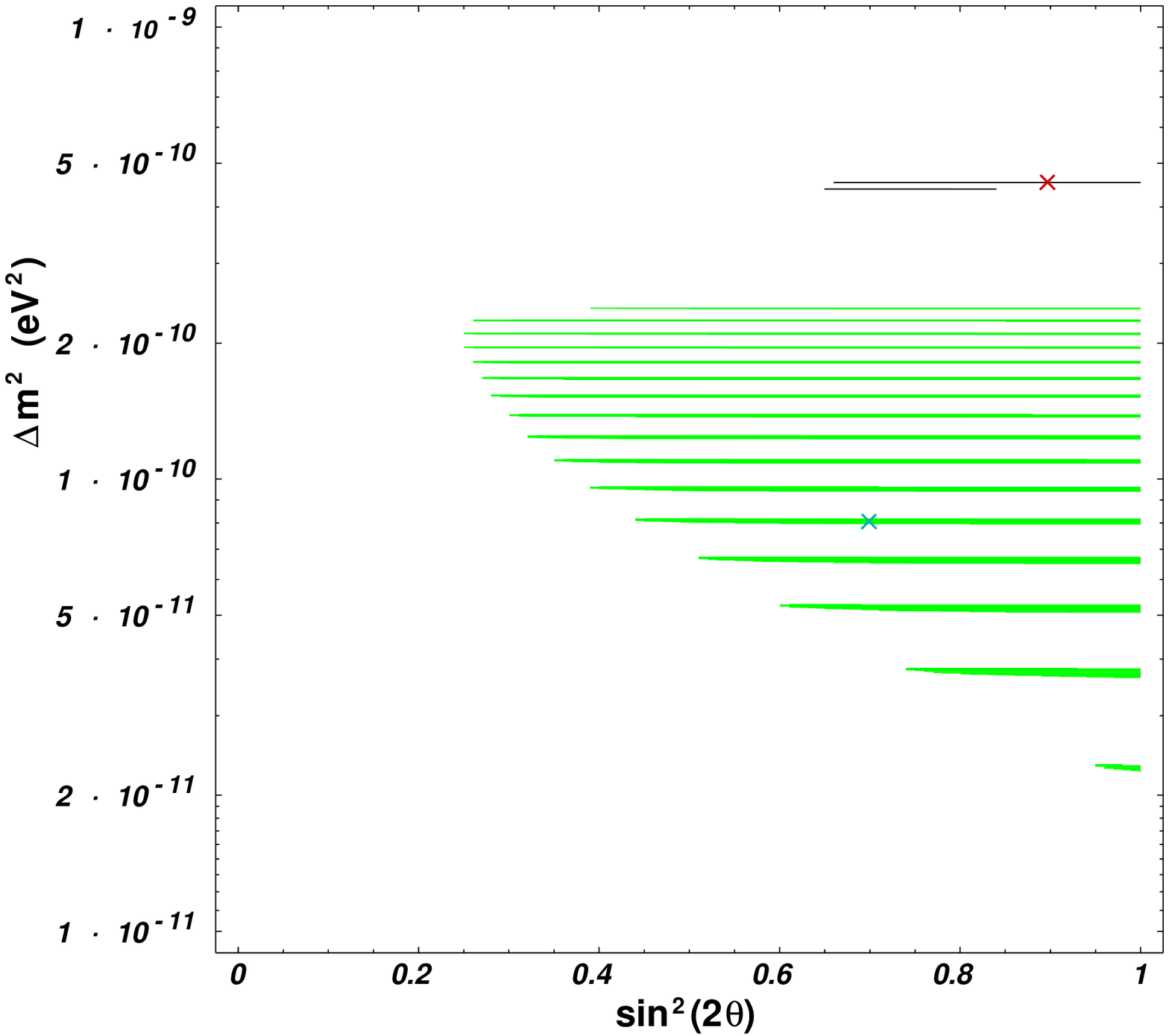,width=0.7\textwidth}
 }
         \caption{Measurement of the neutrino oscillations parameters
   $\sin^22\theta$ and $\Delta m^2$,
   assuming no knowledge of the SSM and the number of background
   events. The regions represent the 95\% confidence level contours, 
   for data consistent with the high (dark) and low points
   (light). The input points are indicated in the figure by the two crosses.   
   See text for details. We assume 3 years of Borexino running.}
\label{measurement}
\end{figure}

In order to measure the oscillation parameters, we
perform a 4 parameter ($s$, $b$, $\sin^22\theta$, and $\Delta m^2$) 
fit to the ``data''. The fit is performed by minimizing $\chi^2$
with respect to the incoming neutrino flux ($s$) and the background
rate ($b$), as in Sec.~\ref{sec:sensitivity}, and computing it for
fixed $\sin^22\theta$ and 
$\Delta m^2$. Fig.~\ref{measurement} depicts the
values of ($\sin^22\theta,\Delta m^2$) and the 95\% CL contours 
(for two degrees of freedom), extracted from the ``data''
consistent with the low (light) and high (dark) points. Note that this
is very different from what was done in the previous section. There,
for each point in the ($\sin^22\theta$,$\Delta m^2$) plane there was a
different ``data'' set, and the ``data'' was fitted by a
non-oscillation theoretical function. Here the ``data'' is fixed
(either the low or the high point), and
is fitted by a theoretical function which assumes neutrino
oscillations.   

One should easily note that the extracted 95\% CL
contour for the high point consists of only two ``islands'', while for the
low point one extracts a collection of ``islands''. The reason for this is
simple. When $\Delta m^2\sim \rm{few}\times10^{-10}$~eV$^2$, the
oscillation length is slightly smaller than
$\Delta L$ (see Eq.~(\ref{DeltaL})). 
This means that the seasonal variation of the ``data'' has a very
particular shape (as one may easily confirm by looking at
Figs.~\ref{fig:illustratepoint}, \ref{points}), which cannot be easily 
mimicked by other values of $\Delta m^2$, even when the background
rate and the incoming flux are varied in the fit procedure.

When $\Delta m^2\sim \rm{several}\times10^{-11}$~eV$^2$, the oscillation
length is larger than $\Delta L$, and the
effect of seasonal variations is less pronounced. There is
a collection of $\Delta m^2$'s that yields the same
qualitative behavior. Because our fit procedure allows for the
background rate and the neutrino flux to float freely, a good agreement
with the ``data'' is met for a large portion of the parameter space.
In order to make this discussion clearer, it is useful to describe
in detail what happens to the number of electron neutrinos reaching
the detector as a function of time. 

In the case of the low point: initially, when the Earth is at the
perihelion, the $\nu_e$ survival probability is small and, as time
progresses, monotonically increases until the Earth reaches the
aphelion (after six
months). The process happens in reverse order in the next six months, as
expected. 
There are many other values of the oscillation length, {\it
i.e.} $\Delta m^2$, 
such that the survival probability monotonically increases for
increasing Earth-Sun
distance and therefore a similar qualitative behavior is to be
expected. The main quantitative difference is in the
ratio of the number of events detected in the perihelion and in the
aphelion, which may be accounted for by varying the background rate and
the incoming neutrino flux. This explains the existence of islands. 
For values of $\Delta m^2$ in between islands, the survival probability either
increases {\it and} decreases for varying Earth-Sun distance, or
monotonically {\it decreases}. The exact location of the islands and
their widths can only be understood by analyzing the fit procedure, in
particular the minimization of $\chi^2$ with respect to the background
rate and the incoming neutrino flux. 
Note that there are no ``islands'' above $\Delta m^2\gtrsim 2.5\times
10^{-10}$~eV$^2$. 
This is because when the oscillation length is small enough
(or $\Delta m^2$ large enough),
the survival probability {\it cannot} only increase for increasing
Earth-Sun distance, but necessarily reaches a maximum before the
aphelion, and then decreases, independent of what the survival
probability at the perihelion is. This situation is qualitatively
different from the low point.

In the case of the high point: initially the survival probability is
close to unity, decreases sharply as the Earth moves further from the Sun,
and then grows rapidly, reaching a maximum when the Earth is close to its
aphelion, because the oscillation length is smaller than
$\Delta L$. In this case, little
variations in the oscillation length,  {\it i.e.} $\Delta m^2$,
produce big qualitative changes, including the position and number of
maxima and minima. There is still a small ambiguity ({\it i.e.}\/ two
``islands'') in determining $\Delta m^2$ for the high point. This happens
when the oscillation length is such that the minimum of the survival
probability happens in March/October and the survival probability is
large enough at the perihelion and the aphelion. The fact that the
absolute values of the number of recoil electrons detected are
different is taken care of by varying the signal and the
background.       

In conclusion, if Nature chose neutrino oscillation parameters such
that $\sin^22\theta$ is large and $\Delta m^2\approx
\rm{few}\times10^{-10}$~eV$^2$, Borexino should be able to measure
these parameters independent of the SSM and any knowledge of the
number of background events, with good precision (especially in
$\Delta m^2$). If $\Delta m^2\approx
\rm{several}\times10^{-11}$~eV$^2$, the determination of oscillation
parameters is not as precise. Better precision can be achieved at
KamLAND, but the ambiguity of solutions in the ``low'' $\Delta m^2$
region still remains.

\setcounter{equation}{0}
\section{Exclusion of Vacuum Oscillations}
\label{sec:exclusion}

In this section, we address the issue of what the experiments can
conclude about vacuum
oscillations if no discrepancy from the normal seasonal variation
effect is detected. In this case, one may be able to measure the
incoming neutrino flux, as outlined in Sec.~\ref{sec:flux}. Two distinct
possibilities will be considered: (1) the measured flux is consistent
with the SSM prediction; (2) the measured flux is suppressed with
respect to the SSM prediction.

In the first case, one would be inclined to trust the SSM prediction 
of the $^{7}$Be neutrino flux and use it in the analysis to exclude vacuum 
oscillations.  This will be discussed in Sec.~\ref{subsec:consistent}.  
On the other hand, in the second case, it is not clear if the reduced 
flux is due to MSW neutrino oscillations, an incorrect SSM prediction 
of the neutrino flux, etc.  This will be discussed in 
Sec.~\ref{subsec:suppressed}.

\subsection{If the Flux is Consistent with the SSM Prediction}
\label{subsec:consistent}
We  simulate ``data'' consistent with the SSM 
and the expected number of background events. The relevant numbers are
quoted in Sec.~\ref{sec:flux}. The ``data'' are
binned into months (12 bins per year), and are illustrated 
in Fig.~\ref{points}, assuming three years of Borexino running. We then fit
to the ``data'' annual distributions that include neutrino oscillations for
a given choice of $(\sin^22\theta,\Delta m^2)$, plus
a constant background. The background rate and the incoming neutrino
flux may be allowed to float in the fit, constrained to a positive number. 

It is important to note that this is
the opposite of what was done in Sec.~\ref{sec:sensitivity}, where the
sensitivity of 
Borexino and KamLAND to vacuum oscillations was studied. There, the
simulated ``data'' were consistent with vacuum oscillations, and one
tried to fit a non-oscillation prediction to the ``data'' by varying
the incoming flux  and/or the background. 
Here, the ``data'' are consistent with no oscillations, and one tries
to fit the ``data'' with a prediction which includes the effect of 
neutrino oscillations
for fixed $(\sin^22\theta,\Delta m^2)$, by varying the incoming
flux  and/or the background. If both the background and the incoming
flux are fixed, {\it i.e.}\/ not allowed to vary in the fit procedure, 
the exclusion and the sensitivity regions are the same. On the other hand,
if both the background rate and the incoming flux are allowed to
float, the exclusion region is expected to be smaller than the
sensitivity region presented in Sec.~\ref{sec:sensitivity}, especially
in the region 
$\Delta m^2\lesssim 10^{-10}$~eV$^2$. This is due to the fact that a large
number of points in the parameter space yield an
annual variation of the $\nu_e$ flux which is much larger than
7\%, but agrees with the shape of the normal seasonal
variation. If in the fit procedure the signal is scaled down to reduce the  
amplitude of the variation and the background scaled up to increase  
the number of events, a good fit to the no oscillation case can be attained.

\begin{figure}[tp]
 \centerline{
   \psfig{file=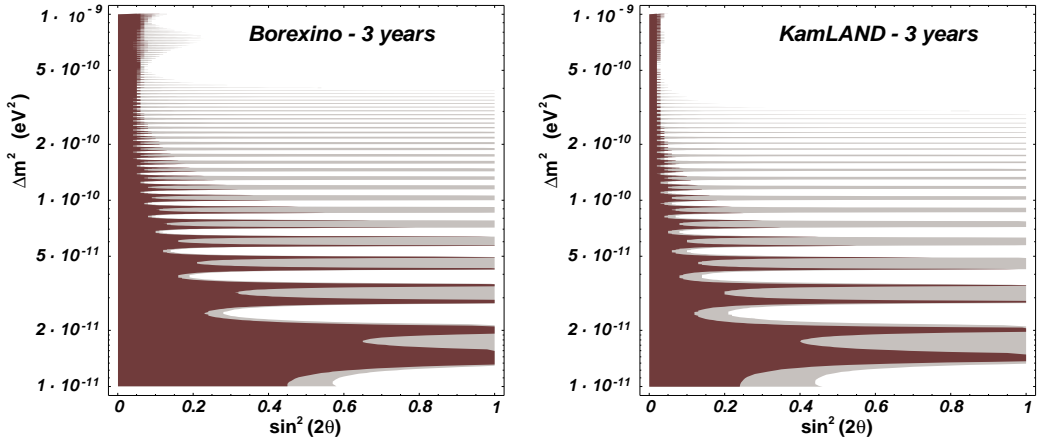,width=1\textwidth}
 }
         \caption{Region of the two neutrino oscillation parameter
   space excluded in the case of no neutrino oscillations if one
   assumes no knowledge of the background and no knowledge of the SSM
   (white) or knowledge of the SSM (light+white), after 3 years of
   Borexino (right) and KamLAND (left) running.}
         \label{exc_b}
\end{figure}

Fig.~\ref{exc_b} shows, for three years of Borexino and KamLAND 
running, the region of the
$(\sin^2 2\theta,\Delta m^2)$ parameter space excluded at 95\% CL,  
if one allows the solar neutrino flux and the
background rate to
float within the positive numbers (in white), and if one
assumes the solar neutrino flux calculated in the SSM within
theoretical errors (in light plus white). 

A few comments are in order. First, one notices that the KamLAND
exclusion region is larger than the one excluded by Borexino. This is,
of course, expected because of KamLAND's larger fiducial volume
and therefore higher statistics.
Second, when the solar neutrino flux is allowed to vary in the fit, 
the excluded region of the parameter space shrinks, as expected and
discussed earlier. 
Third, one can safely claim that, if no discrepancies are detected
in the seasonal variation spectrum, the ``large'' $\Delta m^2$ (several
$\times 10^{-10}$~eV$^2$) set of vacuum
solutions (see Figs.~\ref{fig:finalBorexino} and
\ref{fig:finalKamLAND}) will be excluded, even at
Borexino. Even when no knowledge of
the incoming neutrino flux is used, a reasonable portion
of the ``small'' $\Delta m^2$ (several $\times 10^{-11}$~eV$^2$) set of
solutions is also excluded. When one assumes knowledge of the incoming
neutrino flux, the entire allowed region is excluded.

If the background rate is larger than expected, the excluded region
diminishes accordingly. This is because when the constant background
is enhanced with respect to the oscillation signal it is easier
to achieve a reasonable $\chi^2$ for the fit even when the seasonal
variations due to vacuum oscillations are significantly different from the
no-oscillation case. In particular, when the background rate is large
enough that the seasonal distribution of the data is statistically
consistent with a flat one,
a reasonable $\chi^2$ for the fit can always be achieved simply by
scaling the signal to zero and scaling up the background
appropriately. Explicitly, after three years of Borexino (KamLAND)
running the exclusion region vanishes
if the background rate is 
$\sim$8 (40) times larger than anticipated,
when both
the background rate and the incoming neutrino flux are allowed to
float in the fit or 
$\sim$500 (3000) times larger than anticipated 
when one assumes the neutrino flux predicted by the SSM.  

\subsection{If There is an Overall Suppression of the Flux}
\label{subsec:suppressed}
If there is an overall, {\it i.e.}\/, time-independent suppression of the
flux (which is the case for the MSW solutions), the way to proceed towards
excluding part of the vacuum oscillation parameter space is less
clear. This is because such an experimental result neither agrees with
the SSM prediction nor does it represent any ``smoking gun'' signature
for neutrino oscillations, as is the case of anomalous seasonal
variations. One does not know if the SSM prediction of the flux
is simply wrong, or if there are neutrino oscillations consistent with
one of the MSW solutions or both. Anyway, it is clear that (in general) the
incoming neutrino flux should be considered unknown in the data analysis.

The most conservative option is to follow the same analysis done in the
previous subsection, and allow both the incoming neutrino
flux and the background rate to float in the fit. In this case, the
excluded region of the two-neutrino oscillation parameter space is reduced
significantly, and may completely disappear. This is because
when the number of signal events is reduced the annual distribution is
closer to flat and a good fit is obtained even when the would-be
annual variations are very different. This is very similar to what was
previously discussed at the end of the last subsection, where we
discussed what happens if the background rate turns out to be much larger than
anticipated. Explicitly, after three years of Borexino running and a
signal rate which is 21.3\% of the SSM prediction (as one would obtain
in the case of the small angle MSW solution), Borexino
is unable to exclude any portion of the vacuum
oscillation parameter space, while KamLAND can still exclude about one
half of the ``high'' and ``low'' $\Delta m^2$ preferred regions. 
If the background rate can be estimated by other means with 10\% 
uncertainty, Borexino and KamLAND will be able to exclude the entire``high''
$\Delta m^2$ region and a significant portion of the ``low'' 
$\Delta m^2$ region.

In order to go beyond the most conservative analysis discussed above, 
one would have to look at the overall situation of the solar neutrino 
puzzle at the time of the data analysis.  It is likely that one will be able to do
much better. 
For example, solar neutrino oscillations might have
already been established by the SNO experiment \cite{SNO}, and perhaps
it is reasonable to assume the incoming solar neutrino flux predicted
by the SSM.  
Then it would be possible to exclude a region of the parameter space as large as
the one in Sec.~\ref{subsec:consistent} where one assumes the SSM flux.
Another possibility is that Super-Kamiokande or SNO rules out the
small angle MSW solution by studying the distortions of the 
electron energy spectrum \cite{Super-K_spectrum,SNO}, and a large
suppression of the $^7$Be solar neutrino flux would indicate that there
is something wrong with the SSM.  In this case, it is not clear how to 
proceed.  We do not go into further discussions on all 
logical possibilities.

\setcounter{equation}{0}
\setcounter{footnote}{0}
\section{Conclusions}
\label{sec:conclusion}

We have studied possible uses of the seasonal variation of the $^7$Be
solar neutrino flux at Borexino and KamLAND. Our results can be
summarized as follows.
Once the experiments accumulate enough data to see seasonal
variations, the first step will be to determine if the observed pattern
is consistent with the normal $1/L^2$ flux suppression. If
a discrepancy is found, it will be a sign of vacuum
oscillations. In this case, the seasonal variation of the data can be
used to determine the oscillation parameters $\sin^22\theta$ and
$\Delta m^2$. On the other hand, if the data are consistent with the normal
pattern, the amplitude of the variation can be used to measure the
$^7$Be solar neutrino flux and to exclude a significant portion of
the vacuum oscillation parameter space.

If the observed seasonal variations are consistent with the
normal $1/L^2$ flux suppression, one can use the amplitude of the
variation to determine what fraction of the observed recoil electrons are
induced by the neutrinos coming from the Sun. This method is limited
by statistics, and the accuracy is worse when the $^7$Be solar
neutrino flux is suppressed, as in the case of the small angle MSW
solution. In fact, in Sec.~\ref{sec:flux} we found that in the case of a large
suppression only KamLAND should be able to perform such a measurement,
after 3 years of data taking. It is important to emphasize that
we assumed the oscillation of electron neutrinos into other active flavors. In the
case of oscillations into sterile neutrinos, the $^7$Be solar neutrino
flux might be almost absent, and in this case neither Borexino
nor KamLAND are able to perform a measurement of the flux using this technique.

An important advantage of this technique is that it does not
require a separate estimate of the background rate, which may be a
very difficult task. If the background rate
can be reliably measured by some other means, one can obtain 
another measurement of the neutrino flux. In this case,
the two results can then be
compared for consistency, thus making the final result on the 
$^{7}$Be neutrino flux much more
trustworthy. 

We also studied in great detail the effect of vacuum neutrino
oscillations on seasonal variations.
Our analysis shows that the outlook for discovering vacuum oscillations at
both Borexino and KamLAND is very favorable.
A very important finding in Sec.~\ref{sec:sensitivity} is that the
experiments may detect 
a deviation from
the normal pattern of seasonal variations even without relying on the 
SSM prediction of the incoming neutrino flux or estimate of the
background rate. 
The analysis would consist of trying to fit the observed data with the 
normal $1/L^2$ pattern, treating the incoming neutrino flux and the
background rate as free parameters. With this technique, after three
years of running Borexino should detect anomalous seasonal variations for
almost all values of ($\sin^22\theta,\Delta m^2$) preferred by the
analysis of the neutrino flux data from Homestake, GALLEX, SAGE, and
Super-Kamiokande, as illustrated in
Fig.~\ref{fig:finalBorexino}. The sensitivity region should be
larger at KamLAND (Fig.~\ref{fig:finalKamLAND}). 
Results obtained in this way would be very robust. 
Both experiments are sensitive to an even larger portion of the parameter
space if the background rate can be reliably estimated by auxiliary measurements.

If anomalous seasonal variations are discovered, the data can be 
used to measure the oscillation parameters ($\sin^22\theta$, $\Delta
m^2$). This issue was studied in Sec.~\ref{sec:measurement}. It was
found that for $\Delta 
m^2 \gtrsim 10^{-10}$~eV$^2$ the experiments will be able to determine $\Delta
m^2$ with good precision. At the same time, for
$\Delta m^2 \lesssim 10^{-10}$~eV$^2$ there would be many ``candidate
islands'' in the ($\sin^22\theta,\Delta m^2$) plane, and it will not
be easy to resolve the ambiguity.

On the other hand, the absence of anomalous seasonal variations of the
$^7$Be solar neutrino flux data can be used to exclude regions of the
vacuum oscillation parameter space. In Sec.~\ref{sec:exclusion} we
presented the 
exclusion plots for both Borexino and KamLAND, after three years of
running. An important lesson
from that section is that in order to exclude a large portion of the
preferred region, the
experiments will need to either measure the background rate or rely on
the SSM prediction for the neutrino flux. In the absence of both, the results are
rather weak. This is to be contrasted with the situation in
Sec.~\ref{sec:sensitivity}. 

It is important to keep in mind that the simulated ``data'' is most of 
the time based on the
SSM prediction for the $^7$Be solar neutrino flux and the anticipated
number of background events at Borexino and KamLAND. 
Our {\it numerical} results, therefore, even in the cases when we do not
use the knowledge of the incoming neutrino flux or the background rate 
{\it at the analysis stage}, are not to be regarded as SSM and
background rate independent.  
We would like to draw attention to our comments at the end of
Secs.~\ref{sec:flux} and \ref{sec:sensitivity} on how our results
might change if these 
inputs are changed. We also assume only statistical errors in the data
analysis, neglecting systematic uncertainties due to the lack of
knowledge in the seasonal variation of the background rate. The inclusion of such
effects is beyond the scope of this paper.       

Overall our results indicate that the future Borexino results can
lead to significant progress towards solving the solar neutrino
puzzle. Furthermore, if KamLAND is also able to study solar neutrinos,
one would have access to
a larger data set, and more powerful results can be obtained.
 
\section*{Acknowledgments} 
We would like to express special thanks to Eligio Lisi and Lincoln Wolfenstein
who pointed out the irrelevance of the core size effect.  
HM also thanks John Bahcall, 
Stuart Freedman, and Kevin Lesko for useful conversations.  This work 
was supported in part by the U.S. Department of Energy under Contracts 
DE-AC03-76SF00098, in part by the National Science Foundation under 
grant PHY-95-14797.  HM was also supported by Alfred P. Sloan 
Foundation, and AdG by CNPq (Brazil).

\appendix

\setcounter{equation}{0}
\section{$\chi^2$ Analysis}
\label{explainchi2}

In the analyses in Sec.~\ref{sec:sensitivity}, \ref{sec:measurement},
and \ref{sec:exclusion}, we are interested in the capability of an
``average'' experiment. It is possible to simulate ``data'' with
statistical fluctuations included, but then the value of $\chi^2$ would
vary slightly between different repetitions of the same simulation.
A better approach is to find an expression for $\chi^2$
``averaged'' over many simulations. As we show below, averaging over
statistical fluctuations simply leads to the inclusion of
a constant term in the definition of $\chi^2$.

Suppose we have some solar neutrino data binned into $N_{\rm{bins}}$
bins. Let the average expected
value in the $i$th bin be $d_i$ with corresponding
random fluctuation $\Delta d_i$. Suppose we want to fit this data with a
function $f$, which can depend on two parameters: the signal $s$ and
the 
background $b$. Then the $\chi^2$ of the fit can be defined as follows:
\begin{equation}
\label{defchi2}
 \chi^2(s,b) = \sum_i^{N_{\rm{bins}}} \frac{(d_i+\Delta d_i - f_i(s,b))^2}{\sigma_{d_i}^2},
\end{equation}
where $\sigma_{d_i}=\sqrt{d_i+\Delta d_i}$. Because, in the case of
interest, the
number of events per bin is sufficiently large, we can approximately
set $\sigma_{d_i} \simeq \sqrt{d_i}$.\footnote{One can easily estimate the 
resulting relative error in $\chi^2$ to be of 
$O(1/\sqrt{\left<d_i\right>})$.}

First consider the case when $s$ and $b$ are fixed numbers. 
The average value of the $\chi^2$ one would obtain after simulating the data
many times is
\begin{equation}
\left<\chi^2\right>= 
\left< \sum_i^{N_{\rm{bins}}} \left[\frac{(\Delta d_i)^2}{d_i} + 
\frac{2\Delta d_i(d_i - f_i)}{d_i} +
\frac{(d_i - f_i)^2}{d_i}\right]\right>.
\end{equation}
Using $\left<\Delta d_i\right>=0,\left<(\Delta d_i)^2\right>=d_i$, we find
\begin{equation}
\label{fluct}
\left<\chi^2\right>= 
 \sum_i^{N_{\rm{bins}}} \left[1 + \frac{(d_i - f_i)^2}{d_i}\right]
= {N_{\rm{bins}}} + \sum_i^{N_{\rm{bins}}}  \frac{(d_i - f_i)^2}{d_i}.
\end{equation}
Therefore, in this simplest case it is enough to use the average values $d_i$ 
and the number of bins to compute $\left<\chi^2\right>$.

Next, consider the case when $f(s,b)=b+g(s)$ and $\chi^2$ is minimized
with respect to $b$.
\begin{eqnarray}
\label{minbg}
 \frac{\partial \chi^2(s,b)}{\partial b} &=&
 - \sum_i^{N_{\rm{bins}}} \frac{2 (d_i+\Delta d_i - b - g_i(s))}{d_i} =
 0\nonumber \\
\Longrightarrow b &=& 
   \left(\sum_i^{N_{\rm{bins}}} \frac{(d_i+\Delta d_i - g_i(s))}{d_i}\right)
     \left(\sum_i^{N_{\rm{bins}}} \frac{1}{d_i}\right)^{-1}.
\end{eqnarray}
Introducing $A_i \equiv (d_i+\Delta d_i - g_i(s))/d_i$ and
substituting Eq.~(\ref{minbg}) in Eq.~(\ref{defchi2}), we obtain
\begin{eqnarray}
\label{chi2A}
 \chi^2_{min} &=& \sum_i^{N_{\rm{bins}}} \left[
      \frac{A^2_i}{d_i} - 
      2\frac{A_i}{d_i}\left(\sum_i^{N_{\rm{bins}}}\frac{A_i}{d_i}\right)
      \left(\sum_i^{N_{\rm{bins}}}\frac{1}{d_i}\right)^{-1} +
      \frac{1}{d_i} \left(\sum_i^{N_{\rm{bins}}}\frac{A_i}{d_i}\right)^2
      \left(\sum_i^{N_{\rm{bins}}}\frac{1}{d_i}\right)^{-2}
    \right]\nonumber\\
&=& \sum_i^{N_{\rm{bins}}}\frac{A^2_i}{d_i} - 
    \left(\sum_i^{N_{\rm{bins}}}\frac{A_i}{d_i}\right)^2
     \left(\sum_i^{N_{\rm{bins}}}\frac{1}{d_i}\right)^{-1}.
\end{eqnarray}
Now plugging back in the definition of $A_i$, we perform the averaging
using $\left<\Delta d_i\right>=0,\left<(\Delta d_i)^2\right>=d_i$, and 
$\left<(\Delta d_i)(\Delta d_j)\right>=0$ for $i \neq j$:
\begin{eqnarray}
\label{averages}
 \left<\sum_i^{N_{\rm{bins}}}\frac{A^2_i}{d_i}\right> &=& {N_{\rm{bins}}} + \sum_i^{N_{\rm{bins}}}  \frac{(d_i -
   f_i)^2}{d_i},\\
 \left<\left(\sum_i^{N_{\rm{bins}}}\frac{A_i}{d_i}\right)^2\right>
 &=&\left(\sum_i^{N_{\rm{bins}}} \frac{1}{d_i}\right) +
\left(\sum_i^{N_{\rm{bins}}} \frac{(d_i - g_i(s))}{d_i}\right)^2.
\end{eqnarray}
Substituting Eq.~(\ref{averages}) in Eq.~(\ref{chi2A}), we find
\begin{eqnarray}
 \left<\chi^2_{min}\right> = {N_{\rm{bins}}} - 1 + \sum_i^{N_{\rm{bins}}}  \frac{(d_i -f_i)^2}{d_i} -
 \left(\sum_i^{N_{\rm{bins}}} \frac{(d_i - g_i(s))}{d_i}\right)^2 
     \left(\sum_i^{N_{\rm{bins}}}\frac{1}{d_i}\right)^{-1}. \nonumber \\
\end{eqnarray}
The last two terms are exactly what one would find after minimizing
$\sum_i^{N_{\rm{bins}}} (d_i - b - g_i(s))^2/d_i$ with respect to $b$, and hence in
this case random fluctuations can be accounted for by replacing ${N_{\rm{bins}}}$ in 
Eq.~(\ref{fluct}) by ${N_{\rm{bins}}}-1$.  

One can easily show that, if $f(s,b)_i = b + s\cdot h_i$ and one minimizes
$\chi^2$ with respect to $s$, the effect of random fluctuations is 
also to substitute ${N_{\rm{bins}}}-1$ for ${N_{\rm{bins}}}$ in
Eq.~(\ref{fluct}). The proof is
completely analogous to the case we just studied. 
Moreover, it is straightforward to combine the two
results and consider minimization with respect to both $b$ and $s$, in
which case one should replace ${N_{\rm{bins}}}$ in Eq.~(\ref{fluct})  
by ${N_{\rm{bins}}}-2$.

In general, one should use the number of degrees of freedom
$N_{\rm{d.o.f.}}$ when computing $\left<\chi^2\right>$:
\begin{equation}
\left<\chi^2\right>
= N_{\rm{d.o.f.}} + \sum_i^{N_{\rm{bins}}}  \frac{(d_i - f_i)^2}{d_i}.
\end{equation}

\setcounter{equation}{0}
\section{Analytic Estimate of the Sensitivity Cutoff}
\label{app2}

In Sec.~\ref{sec:sensitivity} we showed that the sensitivity region
for anomalous seasonal 
variations is limited by the finite linewidth of the $^7$Be line.
In this appendix we show how one can analytically estimate the
location and the shape of the sensitivity cutoff.

As was mentioned in
Sec.~\ref{sec:sensitivity}, the true shape of the $^7$Be line is
rather complicated, 
with a Gaussian profile on the low end and an exponential tail
on the high end. For the purpose of this estimate we choose to
approximate the Gaussian part by a sharp cutoff:
\begin{equation}
\label{linepl}
f(E)=\left\{ \begin{array}{ll}
             0 & \textrm{if $E<E_1$}\\
             e^{-a E + b} & \textrm{if $E>E_1$}
           \end{array}. \right.
\end{equation}

To determine the fraction of neutrinos reaching the Earth we integrate 
the oscillation probability $P(E, L)$ given by Eq.~(\ref{osc}) over
the line profile Eq.~(\ref{linepl}) and divide by the normalization
constant $N$.
\begin{eqnarray}
\lefteqn{
        \widetilde{P}(L) =  \frac{1}{N} \int d E P(E, L)f(E) }\nonumber\\ 
&&\simeq\frac{1}{N} \left[\left(1-\frac{\sin^22\theta}{2}\right)
\int_{E_1}^{\infty} d E e^{-a E + b}\right.\nonumber\\
 &&\phantom{\simeq}+\left. \frac{\sin^22\theta}{2}
\int_{E_1}^{\infty} d E \cos\left(2\frac{1.27 \Delta
    m^2 L}{E_0^2}(E_0-E)\right) e^{-a E + b}
\right]\nonumber\\
&&=\frac{e^{-a E_1 + b}}{N} \left[\left(1-\frac{\sin^22\theta}{2}\right)
\frac{1}{a}+ \frac{\sin^22\theta}{2}
\frac{1}{\sqrt{a^2+(1.27 \Delta m^2 L/E_0^2)^2}}
\right.\nonumber\\ 
&&\phantom{=} \left.\times
 \cos\left(2\frac{1.27 \Delta m^2 L}{E_0^2}(E_0-E_1) -
\arctan\left(2\frac{1.27 \Delta m^2 L}{a E_0^2}\right)\right) 
\right].
\end{eqnarray}
Since the width of the line is only several keV while $E_0=0.862$ MeV, 
we can set $E_0-E_1 \simeq E_0$ in the argument of the cosine.
Substituting the value of the normalization constant $N =
\int_{E_1}^{\infty} d E e^{-a E + b}= (1/a) e^{-a E_1 + b}$ and
introducing $\phi \equiv \arctan(2\times1.27 \Delta m^2 L/(a E_0^2))$, 
we obtain
  \begin{eqnarray}
\widetilde{P}(L)\simeq 1
- \frac{\sin^22\theta}{2}\left(1-
\frac{\cos\left(2\frac{1.27 \Delta m^2 L}{E_0} -
\phi\right)}{\sqrt{1+(1.27 \Delta m^2 L/(E_0^2 a))^2}}\right).
\end{eqnarray}

From this equation we can read off the shape of the cutoff. 
Viewed as a function of $\Delta m^2$, for  small values of the mixing
angle the cutoff profile is approximately given by
\begin{equation}
\sin^22\theta_{\rm cutoff} \propto \sqrt{1+(1.27 \Delta m^2 L/(E_0^2 a))^2}.
\end{equation}
Using the numerical value of $a=0.75$~keV$^{-1}$, obtained by fitting
the line profile in \cite{lineprfl}, we find that
$\sin^22\theta_{\rm cutoff}(\Delta m^2)$ should increase by $\sqrt{2}$ with
respect to the smallest value of $\sin^22\theta_{\rm cutoff}$ when
$\Delta m^2 \simeq 2.9 \times 10^{-9}$~eV$^2$. The actual number from
curve 4 in Fig.~\ref{fig:lnwdthandcore} is 
$\Delta m^2 \simeq 1.5 \times 10^{-9}$~eV$^2$. The actual value is
smaller, which is expected, because, for the purpose of this
estimate, we neglected the contribution of the Gaussian part 
of the line profile, effectively making the line narrower.

One can also estimate the location of the cutoff if the line profile
were purely Gaussian (curve 3 in Fig.~\ref{fig:lnwdthandcore}). The
steps are completely analogous: the new normalization constant is 
$N' =\int_{-\infty}^{\infty} d E e^{-(E-E_0)^2/\sigma^2} = 
\sqrt{\pi} \sigma$, and $\widetilde{P}(L)$ is given by
\begin{eqnarray}
\lefteqn{
        \widetilde{P}(L) =  \frac{1}{N'} \int_{-\infty}^{\infty} d E P(E, L) 
e^{-E^2/\sigma^2}}\nonumber\\ 
&&\simeq\frac{1}{N'} \left[\left(1-\frac{\sin^22\theta}{2}\right)
\int_{-\infty}^{\infty} d E e^{-E^2/\sigma^2}\right.\nonumber\\
 &&\phantom{\simeq}+\left. \frac{\sin^22\theta}{2}
\int_{-\infty}^{\infty} d E \cos\left(2\frac{1.27 \Delta
    m^2 L}{E_0^2}(E_0-E)\right) e^{-E^2/\sigma^2}
\right]\nonumber\\
&&= 1-\frac{\sin^22\theta}{2}\left(1 - 
e^{-(1.27 \Delta m^2 L \sigma/E_0^2)^2}\cos\left(2\frac{1.27 \Delta
    m^2 L}{E_0}\right)\right) .
\end{eqnarray}
Thus, the cutoff for this model sets in faster and the profile for
small values of $\sin^2 2\theta$ is Gaussian. Numerically,
$\sin^22\theta_{\rm cutoff}(\Delta m^2)$ is expected to increase by
$\sqrt{2}$ with respect to the smallest value of $\sin^22\theta_{\rm
  cutoff}$ when
$\Delta m^2 \simeq 4.2 \times 10^{-9}$~eV$^2$, which agrees with curve
3 in Fig.~\ref{fig:lnwdthandcore}.

\end{document}